\documentclass[12pt]{iopart}
\usepackage{bm}
\usepackage{graphicx}
\usepackage{color}
\usepackage{subfigure}
\usepackage{hyperref}
\usepackage{latexsym}
\usepackage{amsthm}
\usepackage{amssymb}
\usepackage{braket}
\usepackage{cite}
\usepackage[rgb]{xcolor}
\usepackage{hyperref}

\usepackage{bbold}
\usepackage{bbm}
\usepackage{verbatim}
\usepackage{multirow}
\usepackage[english]{babel}
\usepackage{comment}
\usepackage{bm}
\usepackage{color}
\DeclareGraphicsExtensions{.jpg,.pdf, .mps, .png, .eps, .ps, .EPS,.gif}
\usepackage{url}
\usepackage{multicol}
\usepackage{float}
\usepackage[inline]{enumitem}  
\DeclareMathAlphabet{\mathbcal}{OMS}{cmsy}{b}{n}
\begin{document}

\def\be{\begin{equation}}
\def\ee{\end{equation}}

\def\bc{\begin{center}}
\def\ec{\end{center}}
\def\bea{\begin{eqnarray}}
\def\eea{\end{eqnarray}}
\newcommand{\avg}[1]{\langle{#1}\rangle}
\newcommand{\Avg}[1]{\left\langle{#1}\right\rangle}
\definecolor{applegreen}{rgb}{0.55, 0.71, 0.0}
\def\ie{\textit{i.e.}}
\def\etal{\textit{et al.}}
\def\m{\vec{m}}
\def\G{\mathbcal{G}}


\title[ Hyper-diffusion  on multiplex networks]{Hyper-diffusion on multiplex networks}

\author{Reza Ghorbanchian}

\address{School of Mathematical Sciences, Queen Mary University of London, London, E1 4NS, United Kingdom}
\ead{r.ghorbanchian@qmul.ac.uk}
\author{Vito Latora} 
\address{School of Mathematical Sciences, Queen Mary University of London, London, E1 4NS, United Kingdom}
\address{Department of Physics and Astronomy, University of Catania, and INFN, 95125 Catania, Italy}
\address{Complexity Science Hub Vienna, A-1080 Vienna, Austria}
\ead{v.latora@qmul.ac.uk}
\author{Ginestra Bianconi}
\address{School of Mathematical Sciences, Queen Mary University of London, London, E1 4NS, United Kingdom\\
Alan Turing Institute,The British Library, London NW1 2DB, United Kingdom}
\ead{ginestra.bianconi@gmail.com}
\vspace{10pt}
\vspace{10pt}
\begin{indented}
\item[]
\end{indented}

\begin{abstract}
Multiplex networks describe systems whose interactions can be of different nature, and are fundamental to understand complexity of networks beyond the framework of simple graphs. Recently it has been pointed out that restricting the attention to pairwise interactions is also a limitation, as the vast majority of complex systems include higher-order interactions that strongly affect their dynamics. Here, we propose hyper-diffusion on multiplex networks, a dynamical process in which diffusion on each single layer is coupled with the diffusion in other layers thanks to the presence of higher-order interactions occurring when there exists link overlap. We show that hyper-diffusion on a duplex network (a multiplex network with two layers) can be described by the Hyper-Laplacian in which the strength of four-body interactions among every set of four replica nodes connected by a multilink $(1,1)$ can be tuned by a parameter $\delta_{11}\ge 0$. The Hyper-Laplacian reduces to the standard lower Laplacian, capturing pairwise interactions at the two layers, when $\delta_{11}=0$. By combining tools of spectral graph theory, applied topology and network science we provide a general understanding of hyper-diffusion on duplex networks when $\delta_{11}>0$,  including theoretical bounds on the Fiedler and the largest eigenvalue of Hyper-Laplacians and the asymptotic expansion of their spectrum for $\delta_{11}\ll1$ and $\delta_{11}\gg1$. Although hyper-diffusion on multiplex networks does not imply a direct ``transfer of mass" among the layers (i.e. the average state of replica nodes in each layer is a conserved quantity of the dynamics), we find that the dynamics of the two layers is coupled as  the relaxation to the steady state becomes synchronous when higher-order interactions are taken into account and the Fiedler eigenvalue of the Hyper-Laplacian is not localized in a single layer of the duplex network.
\end{abstract}

\section{Introduction}
Complex systems are characterised by the heterogeneity and complexity of their interactions. Simple pairwise networks can 
only describe systems with two-body interactions, while multiplex networks \cite{bianconi2018multilayer,boccaletti2014structure,kivela2014multilayer} and higher-order networks \cite{battiston2020networks,bianconi2021higher,majhi2022dynamics,torres2021and,battiston2021physics,bick2021higher}
 are necessary to fully represent complex systems where the  interactions can be of different nature and of higher order, i.e.  involving more than two elements at the same time. 
 The research on multiplex and higher-order networks has rapidly expanded in recent years and it has emerged that going beyond simple pairwise interactions significantly enrich our ability to describe the interplay between network structure and dynamics. { Indeed the dynamics on multiplex networks \cite{bianconi2018multilayer} and on higher-order networks \cite{battiston2021physics,majhi2022dynamics,bianconi2021higher}  display a behaviour that is  significantly different  from the corresponding dynamics defined on single pairwise networks. Examples of such notable behaviours are found for different dynamical processes including }  percolation \cite{buldyrev2010catastrophic,sun2022triadic} contagion models \cite{de2017disease,iacopini2019simplicial,ferraz2021phase,st2021universal,st2022influential,higham2021epidemics} evolutionary game theory  \cite{alvarez2021evolutionary},  synchronisation \cite{millan2020explosive,ghorbanchian2021higher,gambuzza2021stability,skardal2019abrupt,del2016synchronization,salova2021cluster,zhang2021unified,jalan2016cluster,chutani2021hysteresis} and diffusion models \cite{gomez2013diffusion,sole2013spectral,de2014navigability,radicchi2013abrupt,jost2019hypergraph,mulas2020coupled,mulas2022graphs,sahasrabuddhe2021modelling,neuhauser2020multibody,carletti2020random,torres2020simplicial,taylor2022,millan2021local,reitz2020higher,bianconi2021topological,krishnagopal2021spectral}.
 {However until now, if we exclude few recent works \cite{ferraz2021phase,sun2021higher,kim2021link}, most of the literature  has considered only multilayer network dynamics}  {with strictly pairwise interactions.} 

Here, we introduce and study a novel type of diffusion process on multiplex networks, that we name {\em hyper-diffusion}. 
 {In hyper-diffusion  the two layers 
of a duplex network (a multiplex network with two layers) are  coupled by higher-order interactions occurring when  the links in the two layers overlap.}

Diffusion on multiplex networks has been widely discussed in the literature, and proven to have a significant role in modelling real systems. Diffusion on multiplex networks is typically described by a theoretical framework proposed in Refs.\cite{gomez2013diffusion,sole2013spectral,de2014navigability,radicchi2013abrupt} that is inspired by transportation networks. In the transportation network of a city there are different modes of transportation, e.g. bus, underground, etc., and it is in general possible to switch from one mode to another by paying an additional cost.  {Motivated by this example,} in Refs. \cite{gomez2013diffusion,sole2013spectral,de2014navigability} it is assumed that diffusion occurring  {on} each layer of the multiplex network is coupled by interlinks, i.e. connections existing among replica nodes of the multiplex network,  {which allows to diffuse from one layer to another.} For instance, in the transportation network of London an interlink can connect Oxford Circus tube station to Oxford Circus bus station. However, not every multiplex network might be suitable for this framework, as it is not always possible to give a physical meaning to interlinks.

Diffusion on higher-order networks is also attracting large attention with recent works related to consensus models \cite{sahasrabuddhe2021modelling,neuhauser2020multibody}, random walks on higher-order networks \cite{carletti2020random}, diffusion and higher-order diffusion on simplicial complexes \cite{torres2020simplicial,taylor2022,millan2021local,reitz2020higher,bianconi2021topological},  diffusion on hypergraphs and  {on} oriented  {hypergraphs} \cite{jost2019hypergraph,mulas2020coupled,mulas2022graphs,ferraz2021phase}.
Of special interest for this work is the recent use of applied topology and spectral graph theory to treat diffusion on oriented hypergraphs that has been recently proposed and investigated in Refs. \cite{jost2019hypergraph,mulas2020coupled,mulas2022graphs,jost2021normalized}. Oriented hypergraphs are formed by hyperedges whose interacting elements, nodes of the hyperedge, are distinguished between input and output nodes; similarly to what happens in reaction networks that constitute a major application of this framework. In this context, the authors of Refs. \cite{jost2019hypergraph,mulas2020coupled,jost2021normalized,mulas2022graphs} have proposed a Laplacian matrix that can be used to investigate diffusion, as well as synchronisation \cite{gallo2022synchro} in oriented hypergraphs. 

Empirical investigation of the structure of multiplex networks  \cite{bianconi2013statistical,bianconi2018multilayer,menichetti2014weighted,bentley2016multilayer,battiston2017multilayer} has shown that the presence of link overlap between the layers of multiplex networks is a ubiquitous property of real systems and can significantly affect their dynamical properties. For instance, very often two 
airports are connected by flights operated by different airlines, while two scientists are often connected both in the citation and in the collaboration network. 
The link overlap among the layers of a multiplex network can  be captured by the use of multilinks \cite{bianconi2013statistical,bianconi2018multilayer} characterising for each pair of nodes of the multiplex network, the set of layers in which the two nodes are linked.  {For instance, in a duplex network, there are four possible types of multilinks: two nodes are connected by  
a multilink of type $(1,0)$ or type $(0,1)$ if there is a link respectively at the first layer or at the second layer only,  
by a multilink $(1,1)$ if they are connected in both layers, and finally by 
multilink $(0,0)$ if there are no links at both layers.} 
In the hyper-diffusion process on duplex networks that we propose in this paper, the coupling between the diffusion dynamics taking place in  the two layers is provided by the higher-order interactions encoded by multilinks of type $(1,1)$, i.e.~by the $4$-body interactions existing among the $4$ replica nodes connected by two overlapping links. 
We will motivate our definition of hyper-diffusion with arguments coming from applied topology, and we will show that hyper-diffusion is driven by  Hyper-Laplacians of multiplex networks that are inspired by the Laplacians of hypergraphs introduced in Ref.\cite{jost2019hypergraph,mulas2020coupled}.
In our work, we will investigate the properties of hyper-diffusion using spectral graph theory \cite{chung1997spectral} and asymptotic expansions of the eigenvalues, whose results will be confirmed by numerical simulations.
We will provide analytical upper and lower bounds for the Fiedler eigenvalues of both un-normalised and normalised Hyper-Laplacian and we will relate this spectral properties with the dynamics of hyper-diffusion.
Although hyper-diffusion does not involve a ``transfer of mass" among the layers of the duplex network, as the average dynamical state of the replica nodes of each single layer is a conserved quantity of the dynamics, we will show that in hyper-diffusion the relaxation time to the steady state is the same in every layer of the multiplex network, as long as the Fiedler eigevector of the Hyper-Laplacian is not localized on a single layer.

The paper is organised as follows. In Sec. 2 we introduce duplex networks and the Laplacian matrices for each of their layers. In Sec. 3 we introduce multilinks and multi-Laplacian matrices. In Sec. 4 we define the hyper-boundary operators of a duplex network encoding for pairwise and higher-order interactions. In Sec. 5 we introduce the Hyper-Laplacians  {of a duplex network, the 
central operators of our work}, 
while in Sec. 6 we characterise their major spectral properties. In Sec. 7 we introduce and investigate hyper-diffusion on duplex networks. In Sec. 8 we focus on the asymptotic behaviour of hyper-diffusion in the limiting case in which higher-order  {diffusion is} a perturbation to the diffusion dynamics in different layers and in the limiting case in which it is instead the dominant contribution to hyper-diffusion. Finally, in Sec. 9 we provide the concluding remarks. The paper is enriched by extensive set of Appendices providing the necessary proofs for the statements made in the main body of the paper.

\begin{figure}[!htbp]
\begin{center}
\includegraphics[width=0.5\textwidth]{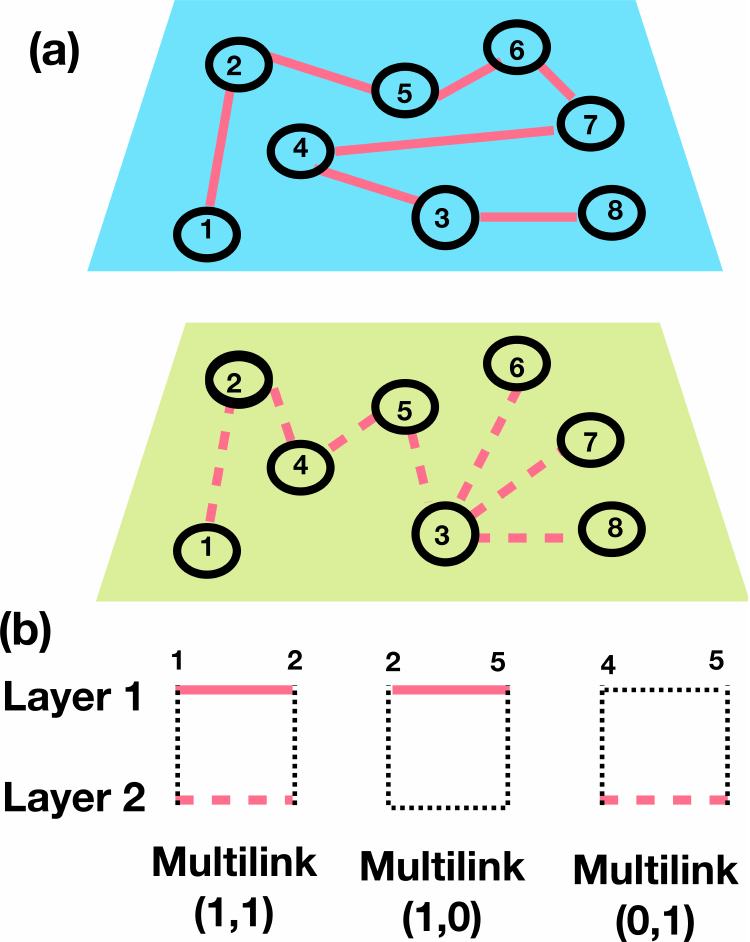}
\caption{{\bf Schematic representation of a duplex network.} We consider a duplex network of $M=2$ layers and $N=8$ nodes (panel (a)).  {The three different types of non-trivial multilinks are shown in panel (b): multilink  of type $(1,1)$ ,  $(1,0)$, and $(0,1)$ indicating for instance the connectivity among nodes nodes $2$ and $5$;  nodes $4$ and $5$; and nodes $1$ and $2$ respectively.}}
\label{fig:1}
\end{center}
\end{figure}
\section{Duplex networks and their two layers}
A duplex network $\vec{G}=(G^{[1]},G^{[2]})$ is a multiplex network formed by $M=2$  layers, $G^{[\alpha]}=(V^{[\alpha]},E^{[\alpha]})$ with $\alpha\in \{1,2\}$, each one formed by a set $V^{[\alpha]}$ of $N=|V^{[\alpha]}|$ replica nodes $v_i^{[\alpha]}$ with  $i\in \{1,2,...,N\}$ and a set of  links $E^{[\alpha]}$ of cardinality $K^{[\alpha]}=|E^{[\alpha]}|$ (see Figure  $\ref{fig:1}$). Here, we indicate the set of all replica nodes of the duplex network by $V=V^{[1]}\cup V^{[2]}$, with $|V|=2N$.
Each layer $G^{[\alpha]}$ of the network can be fully described by its $N\times N$ adjacency matrix $a^{[\alpha]}$. The degree of the node $v_i^{[\alpha]}$ in layer $\alpha$ is indicated by $k_i^{[\alpha]}$ with $k_{i}^{[\alpha]}=\sum_{j=1}^Na_{ij}^{[\alpha]}$.
In a duplex network, replica nodes belonging to different layers are paired so that  $(v_i^{[1]}$, $v_i^{[2]})$  represents the state of a  given node $v_i$ in layer $1$ and $2$ respectively. We indicate the set of all nodes of the duplex network by $\hat{V}$ with $|\hat{V}|=N$.

In this work we are interested in treating duplex networks with tools of algebraic topology and spectral graph theory in order to describe  hyper-diffusion, i.e.~a diffusion process that takes into account higher-order interactions between replica nodes. In order to set the ground for this investigation, we start by defining  the boundary operators of each layer $\alpha$. To this end, we introduce a notion of orientation of the links.
In particular  {we associate each link  in layer $\alpha$ with an orientation induced by the node labels, i.e. $\ell^{[\alpha]}=(v_i^{[\alpha]}, v_j^{[\alpha]})$ has positive orientation if and only if  $i<j$. We indicate the $k$-th positively oriented link in layer $\alpha$ as $\ell_k^{[\alpha]}$.} 

 {A $0$-cochain is a generic  function $f^{[\alpha]}$ defined on the nodes of a given layer $\alpha$, while a $1$-cochain is a generic function $g^{[\alpha]}$ defined on oriented edges and such that 
\bea
g^{[\alpha]}(v_i^{[\alpha]},v_j^{[\alpha]})=-g^{[\alpha]}(v_j^{[\alpha},v_i^{[\alpha}).
\eea}
 {  The  boundary operator $\nabla_{\alpha}$ of layer $\alpha$ maps $0$-cochains to $1$-cochains. Given a $0$-cochain $f^{[\alpha]}$ and a positively  oriented link $\ell_k^{[\alpha]}=(v_{i}^{[\alpha]}, v_j^{[\alpha]})$, the boundary operator $\nabla_{\alpha}$}  
 {is defined as:}
\bea
\nabla_{\alpha} f^{[\alpha]}\ (\ell^{[\alpha]}_k    
= (v_{i}^{[\alpha]}, v_j^{[\alpha]})   )= 
\begin{array}{ll}
       f^{[\alpha]}(v_i^{[\alpha]})-f^{[\alpha]}(v_j^{[\alpha]}),
    \end{array}
\eea
 {with $\nabla_{\alpha} f^{[\alpha]}\ (-\ell^{[\alpha]}_k)=-\nabla_{\alpha} f^{[\alpha]}\ (\ell^{[\alpha]}_k)$.\\
The co-boundary operator $\nabla^*_{\alpha}$ of layer $\alpha$ maps $1$-cochains to $0$-cochains.
Given a $1$-cochain  $g^{[\alpha]}$ defined on the links of a given layer $\alpha$, the co-boundary operator $\nabla^*_{\alpha}$ of layer $\alpha$ 
 {is defined as:}
\bea
\nabla^*_{\alpha} g^{[\alpha]}\ (v^{[\alpha]}_i)=\begin{array}{ll}
    \sum_{\ell^{[\alpha]}_k=(v_i,v_j)\in E^{[\alpha]}}g(\ell_k^{[\alpha]})-\sum_{ \ell^{[\alpha]}_k=(v_j,v_i)\in E^{[\alpha]}}g(\ell_k^{[\alpha]}). 
    \end{array}
    \eea}

The co-boundary operator defined above can be represented by the $N\times K^{[\alpha]}$ incidence matrix $B^{[\alpha]}$ of elements
\bea
B^{[\alpha]}_{ik}=\left\{\begin{array}{ll}
1&\mbox{if}\ \ell^{[\alpha]}_{k}=(v_i^{[\alpha]},v_j^{[\alpha]}),\nonumber \\
-1&\mbox{if}  \ \ell_{k}^{[\alpha]} =(v_j^{[\alpha]},v_i^{[\alpha]}).
\end{array}\right.
\eea
 Note that the boundary and co-boundary operators defined above are adjoint of each other  {with respect to the standard $L^2$ inner product.}\\
Given the  {$0$-cochain} $f^{[\alpha]}$ defined on the nodes in layer $\alpha$, the Laplace operator $\Delta_{\alpha}$ for layer $\alpha$ can be defined as: 
\be
\Delta_{\alpha} f\ (v_i^{[\alpha]})=\nabla_{\alpha}^*(\nabla_{\alpha} f)(v_i^{[\alpha]})=\sum_{l_k^{[\alpha]}=(v_i^{[\alpha]},v_j^{[\alpha]})}f^{[\alpha]}(v_i^{[\alpha]})-f^{[\alpha]}(v_j^{[\alpha]}).
\ee
The matrix corresponding to the Laplace operator $\Delta_{\alpha}$ is the so-called Laplacian matrix $L^{[\alpha]}$ given by:
\be
L^{[\alpha]}=B^{[\alpha]}(B^{[\alpha]})^\top,
\ee
whose elements are  $L^{[\alpha]}_{ij}=k_i^{[\alpha]}\delta_{ij}-a^{[\alpha]}_{ij}$.

By introducing diagonal matrices $D^{[\alpha]}$, 
whose diagonal elements are given by the node degrees in layer $\alpha$, i.e. $D^{[\alpha]}_{ii}=k_i^{[\alpha]}$, we can define the normalised (symmetric) Laplacian of layer $\alpha$ in the usual way as:    
\bea
L^{[\alpha]}_{norm}=(D^{[\alpha]})^{-\frac{1}{2}}L^{[\alpha]}(D^{[\alpha]})^{-\frac{1}{2}},
\eea
where we have adopted the convention that $(D^{[\alpha]})^{-1}_{ii}=0$ if $k_i^{[\alpha]}=0$.

The two matrices $L^{[\alpha]}$ and $L^{[\alpha]}_{norm}$ obeys the usual spectral properties of un-normalised and normalised graph Laplacians  \cite{chung1997spectral}. In particular $L^{[\alpha]}$ and $L^{[\alpha]}_{norm}$ are semi-definite positive with a zero eigenvalue of multiplicity equal to the number of connected component of the network in layer $\alpha$. Moreover, the normalised Laplacian $L^{[\alpha]}_{norm}$ has eigenvalues smaller or equal to $2$.

 \section{Multilinks and multiboundary operators }
In a single network every two nodes can be either linked or not linked. In a multiplex network, instead any two nodes can be connected in multiple ways.  Multilinks \cite{bianconi2013statistical,bianconi2018multilayer} (see Figure  $\ref{fig:1}$) characterise exhaustively the pattern of connection between any two nodes in a multiplex network. Specifically, in a duplex network, any two nodes $v_i$ and $v_j$ are connected by a multilink of type $\vec{m}=\big(m^{[1]}, m^{[2]}\big)$ with elements $m^{[\alpha]}\in\{1,0\}$,  {where $m^{[\alpha]}=1$ iff there is a link between the two nodes at layer $\alpha$, while $m^{[\alpha]}=0$ otherwise.}
%
%
Therefore, there are four possible types of multilinks. If nodes $v_i$ and $v_j$ are linked in both layers we say that $v_i$ and $v_j$  are connected by a multilink of type $(1,1)$. If the nodes $v_i$ and $v_j$ are linked by a link only in the  first layer but they are not linked in the second layer, we say that they are connected by a  multilink of type $(1,0)$. Conversely, if nodes $v_i$ and $v_j$ are linked only in  {the second} layer, 
we say they are connected by a multilink of type $(0,1)$. Finally two nodes are connected by the  trivial multilink of type $(0,0)$ when there is no link between the two nodes in neither layer. 
For each type of multilink $\vec{m}$, we can define a corresponding $N \times N$ multiadjacency matrix $A^{\vec{m}}$ having elements \cite{bianconi2018multilayer}:  
\bea
A^{\vec{m}}_{ij}=\left\{\begin{array}{lc} 1 & \mbox{if}\  \vec{m}_{ij}=\vec{m},\\0 & \mbox{otherwise}. \end{array}\right.
\eea
Therefore, $A^{\vec{m}}_{ij}=1$  if there is a multilink $\vec{m}$ between nodes $v_i$ and $v_j$,  otherwise $A^{\vec{m}}_{ij}=0$.  {Please note however that
since any two nodes $v_i$ and $v_j$ can be  connected at most by a  single multilink,   the entries of the multiadjacency matrices are not independent \cite{bianconi2013statistical}.}
This  implies, for example, that if  two nodes $v_i$ and $v_j$ are connected only in layer one, then $A^{(1,0)}_{ij}=1$ and the corresponding elements of the other multi-adjacency matrices are  {all} $0$, i.e. $A^{(0,1)}_{ij}=A^{(1,1)}_{ij}=A^{(0,0)}_{ij}=0$.\\
Given the multi-adjacency matrices it is possible to define the multidegree\cite{bianconi2013statistical,bianconi2018multilayer} $k^{\vec{m}}_i$ of  node $v_i$  as the number of multilinks of type $\vec{m}$ incident to node $v_i$, i.e.
\begin{equation}
k^{\vec{m}}_i=\sum_{j=1}^{N} A^{\vec{m}}_{ij}.
\end{equation}
Clearly the degrees $k_i^{[\alpha]}$ are related to the multidegrees of node $i$ by
\bea
k^{[1]}_i=k_i^{(1,0)}+k_i^{(1,1)},\quad k^{[2]}_i=k_i^{(0,1)}+k_i^{(1,1)}.
\eea
Interestingly multilinks lend themselves to be  studied also from the applied topology perspective. For this purpose, it is possible to define multi-incidence matrices $B^{\vec{m}}$, which extend the notion of incidence matrix of single networks to the case of multiplex networks 
\cite{dirac}.  
In particular they  represent the co-boundary operator acting on multilinks  of type $\vec{m}$. In order to define multi-incidence matrices, we consider oriented multilinks, in which we  adopt the orientation induced by the node labels, although the node labels can be chosen arbitrarily.  {A multilink between node $v_i$ and $v_j$, i.e. $\ell^{\vec{m}}=(v_i,v_j)$ will have  positive orientation if  $i<j$ and negative orientation otherwise. We will indicate with $\ell_k^{\vec{m}}$ the $k-th$ positively oriented multilink.}   {Consider $0$-cochain $f$ defined on the  nodes $v_i\in \hat{V}$} of a multiplex network, the multi-boundary operator ${\nabla}_{\vec{m}}$ 
is defined as
\be
{\nabla }_{\vec{m}}f (\ell_k^{\vec{m}}=(v_i,v_j))=f(v_i)-f(v_j),
\ee
 {with 
${\nabla }_{\vec{m}}f (v_j,v_i)={\nabla }_{\vec{m}}f (-(v_i,v_j))=-{\nabla }_{\vec{m}}f (v_i,v_j).$
}

 {The adjoint operator of ${\nabla}_{\vec{m}}$ (always with respect to the $L^2$ inner product), is co-boundary operator ${\nabla}^*_{\vec{m}}$. Let  us consider the $1$-cochain $g^{\vec{m}}$ defined on the multilinks of type $\vec{m}$, the  co-boundary operator ${\nabla}^*_{\vec{m}}$ %
is defined as}
\be
{\nabla}^*_{\vec{m}}g^{\vec{m}}(v_i)=\sum_{\ell_k^{\vec{m}}=(v_i,v_j)}g(\ell_k^{\vec{m}})-\sum_{\ell_k^{\vec{m }}=(v_j,v_i)}g(\ell_k^{\vec{m}}).
\ee
The matrix corresponding to the co-boundary operator ${\nabla}^*_{\vec{m}}$ is the $N\times K^{\vec{m}}$ multi-incidence matrix $B^{\vec{m}}$ whose elements are given by  
\bea
B^{\vec{m}}_{ik}=\left\{\begin{array}{ll}1 & \mbox{if} \ {\ell}_k^{\vec{m}}= (v_i,v_j), \\-1 &\mbox{if}\ {\ell}_k^{\vec{m}}=(v_j, v_i),\\0 &\mbox{otherwise}.\end{array}\right.
\eea
As an example, in a two-layer multiplex network, we have three non-trivial multi-incidence matrices $B^{(1,0)}$, $B^{(0,1)}$, $B^{(1,1)}$ and one trivial multi-incidence matrix $B^{(0,0)}$. \\
Given the $0$-cochain $f:\hat{V} \rightarrow \mathbb{R}$ defined on the nodes of the multiplex network, the Laplace operator $\Delta_{\vec{m}}$ for multilink $\vec{m}$ is defined as
\be
(\Delta_{\vec{m}}f)(v_i)=\nabla^*_{\vec{m}}(\nabla_{\vec{m}}f)(v_i)=\sum_{\ell_k^{\vec{m}}=(v_i,v_j)}f(v_i)-f(v_j),
\ee
which is represented by $N \times N$ multi-Laplacian matrix $L^{\vec{m}}$ which reads
\begin{equation}
    L^{\vec{m}}_{ij}=k_i^{\vec{m}}\delta_{ij}-A^{\vec{m}_{ij}}.
\end{equation}
By introducing the diagonal matrices $D^{\vec{m}}$, having the multidegrees as diagonal elements, i.e. $D^{\vec{m}}_{ii}=k_i^{\vec{m}}$, we can also introduce the normalised multi-Laplacian
\bea
L^{\vec{m}}_{norm}=(D^{\vec{m}})^{-\frac{1}{2}}L^{\vec{m}}(D^{\vec{m}})^{-\frac{1}{2}},
\eea
where we have adopted the convention that $(D^{\vec{m}})^{-1}_{ii}=0$ if $k_i^{\vec{m}}=0$.
The properties of each multi-Laplacian $L^{\vec{m}}$ and each normalized multi-Laplacian $L^{\vec{m}}_{norm}$ are the usual properties of the graph Laplacians defined on a single network, however the different multi-Laplacians of the same duplex networks   might display non-trivial relations that can be probed by  {evaluating} 
their commutators \cite{dirac}. 

\section{Hyper-boundary operators of a duplex network }

In this section we introduce the hyper-boundary operators of a duplex networks, indicated also as lower and higher boundary operators.
These definitions are inspired by the definition of boundary operators recently proposed for hypergraphs \cite{jost2019hypergraph,mulas2020coupled,mulas2022graphs}.
The hyper-boundary operators are defined on a duplex network in which we treat each node $v_i$ as a direct sum of their replica nodes $v_i^{[1]}$ and $v_i^{[2]}$ and we assume that each replica node can be in a different dynamical state. As we will see in the following,  the  lower boundary operator captures the pairwise interactions between linked replica nodes belonging to the same layer, while the higher boundary operator captures the four-body interactions existing among  every set of four replica nodes connected by a multilink of type $(1,1)$.  {In particular, the oriented multilinks of type  $(1,1)$ are here treated similarly as  oriented hyperedges formed by two input and two output nodes as in Refs.\cite{jost2019hypergraph,mulas2020coupled,mulas2022graphs}. }

 {Consider a $0$-cochain $f:V\rightarrow \mathbb{R}$ defined on the $2N$ replica  nodes of a duplex network.}   {The value of the $0$-cochain $f$  on the generic node $v_i^{[1]}$ of the first layer is indicated as  $f^{[1]}(v_i^{[1]})$, while the value that the $0$-cochain $f$ takes on the generic node $v_i^{[2]}$ of the second layer is indicated as $f^{[2]}(v_i^{[2]})$. }
The lower-order boundary operator $\nabla_{lower} $ maps a generic $0$-cochain to a $1$-cochain and is defined by 
\bea
\hspace*{-20mm}\nabla_{lower} f\ (\ell_k^{[\alpha]})=\left\{\begin{array}{ll}
       \nabla_{1} f^{[1]}\ (\ell_k^{[1]})=f^{[1]}(v_i^{[1]})-f^{[1]}(v_j^{[1]}) & \mbox{if}\  \alpha=1, \  \ell^{[\alpha]}_k=(v_i^{[1]},v_j^{[1]}),
       \\[.2cm]
       \nabla_{2} f^{[2]}\ (\ell_k^{[2]})=f^{[2]}(v_i^{[2]})-f^{[2]}(v_j^{[2]})  & \mbox{if}\ \alpha=2,\ \ell^{[\alpha]}_k=(v_i^{[2]},v_j^{[2]}),
    \end{array}\right.
\eea
 {with $\nabla_{lower} f\ (-\ell_k^{[\alpha]})=-\nabla_{lower} f\ (\ell_k^{[\alpha]}).$ \\
Let us consider a $1$-cochain $g$ defined on the oriented links belonging to both layers of the  duplex network.}
The lower-order co-boundary operator $\nabla^*_{lower} $ maps $1$-cochains $g$ to $0$-cochains and is defined by 
\bea
    \hspace*{-20mm}\nabla^*_{lower} g\ (v^{[1]}_i)&=\nabla^*_1g\ (v_i^{[1]})=&\sum_{ \ell^{[1]}_k=(v_i^{[1]},v_j^{[1]}) }g(\ell_k^{[1]})-\sum_{  \ell^{[1]}_k=(v_j^{[1]},v_i^{[1]})\in E^{[1]}}g(\ell_k^{[1]}), \nonumber \\[.2cm]
     \hspace*{-20mm}\nabla^*_{lower} g\ (v^{[2]}_i)&=\nabla^*_2g\ (v_i^{[2]})=&\sum_{ {\ell}^{[2]}_k=(v_i^{[2]},v_j^{[2]})}g(\ell_k^{[2]})-\sum_{  \ell^{[2]}_k=(v_j^{[2]},v_i^{[2]})}g(\ell_k^{[2]}).\nonumber 
\eea
The lower order Laplace operator $\Delta^{lower}$ of a duplex network can therefore be defined as
\be
\Delta^{lower}=\nabla^*_{lower}(\nabla_{lower} f)(v),
\ee
which reads
\bea
 \Delta^{lower} f\ {(v_i^{[1]})}&=\sum_{\ell_k^{[1]}=(v_i^{[1]},v_j^{[1]})} f^{[1]}(v_i^{[1]})-f^{[1]}(v_j^{[1]}),\nonumber \\
  \Delta^{lower} f\ {(v_i^{[2]})}&=\sum_{\ell_k^{[2]}=(v_i^{[2]},v_j^{[2]})} f^{[2]}(v_i^{[2]})-f^{[2]}(v_j^{[2]}).\nonumber 
 \eea
The matrix $\mathcal{L}^{lower}$ corresponding to the Laplace operator $\Delta^{lower}$ is the direct sum of $L^{[1]}$ and $L^{[2]}$, namely we have: 
\be
\mathcal{L}^{lower}=L^{[1]}\oplus L^{[2]}=\left(\begin{array}{cc}
    L^{[1]} & {0} \\
    {0} & L^{[2]}
\end{array}\right). 
\ee
Therefore, the operator $\Delta^{lower}$ and its corresponding matrix $\mathcal{L}^{lower}$ capture  pairwise interactions between adjacent nodes in the two layers of the duplex network.
Since $\mathcal{L}^{lower}$ is the direct sum of $L^{[1]}$ and $L^{[2]}$,  the spectrum of $\mathcal{L}^{lower}$ is formed by the concatenation of the eigenvalues of $L^{[1]}$ and $L^{[2]}$. It follows that the zero eigenvalue has degeneracy given by the sum of the 
 {numbers of} 
connected components of 
 {the two} 
layers, and that the Fiedler eigenvalue (the smallest non zero eigenvalue)  $\lambda_{s}(\mathcal{L}^{lower})$  of $\mathcal{L}^{lower}$ is given by the minimum between the Fiedler eigenvalues of $L^{[1]}$ and $L^{[2]}$, i.e.
\bea
\lambda_s(\mathcal{L}^{lower})=\min\left(\lambda_s(L^{[1]}),\lambda_s(L^{[2]})\right).
\label{lower_l2}
\eea
In order to define  {the} higher-order operators on  {a} duplex network, we make use of multilinks of type $\vec{m}=(1,1)$.  {The higher-order boundary operator $\nabla_{higher}$ maps a $0$-cochain $f$ defined on the $2N$ replica nodes to a $1$-cochain defined on multilinks of type $\vec{m}=(1,1)$. Indicating by  $\ell_k^{(1,1)}$  the $k$-th positively oriented multilink of type $\vec{m}=(1,1)$ connecting nodes $v_i$ and $v_j$, (i.e. $\ell_k^{(1,1)}=({ v}_i,{ v}_j)$ with $i<j$) the higher-order boundary operator $\nabla_{higher}$ is  defined by }  
\bea
\hspace*{-25mm}\nabla_{higher} f\ (\ell_k^{(1,1)})&=&
   \nabla_{(1,1)}f^{[1]}(\ell_k^{(1,1)})+\nabla_{(1,1)}f^{[2]}(\ell_k^{(1,1)})\nonumber  \\
   &=& f^{[1]}(v_i^{[1]})-f^{[1]}(v_j^{[1]})+f^{[2]}(v_i^{[2]})-f^{[2]}(v_j^{[2]}),
\eea
 {where  $\nabla_{higher} f\ (-\ell_k^{(1,1)})=-\nabla_{higher} f\ (\ell_k^{(1,1)}).$\\
The higher-order co-boundary operator $\nabla^*_{higher}$ is the adjoint operator  of $\nabla_{higher}$ with respect to the standard $L^2$ inner product and it is defined as}
\bea
    \hspace*{-20mm}\nabla^*_{higher} g\ (v^{[1]}_i)&=\nabla^*_{(1,1)}g\ (v_i^{[1]})=&\sum_{ \ell^{(1,1)}_k=({v}_i,v_j)}g(\ell_k^{(1,1)})-\sum_{  \ell^{(1,1)}_k=(v_j,v_i)}g(\ell_k^{(1,1)}), \nonumber \\[.2cm]
     \hspace*{-20mm}\nabla^*_{higher} g\ (v^{[2]}_i)&=\nabla^*_{(1,1)}g\ (v_i^{[2]})=&\sum_{ \ell^{(1,1)}_k=(v_i,v_j)} g(\ell_k^{(1,1)})-\sum_{  \ell^{(1,1)}_k=(v_j,v_i) }g(\ell_k^{(1,1)}). \nonumber
\eea
Consequently, the higher-order Laplace operator $\Delta^{higher}$ of a duplex network is defined as
\be
\Delta^{higher}=\nabla^*_{higher}(\nabla_{higher} f)(v),
\ee
which reads
\bea
 &(\Delta^{higher} f){(v_i^{[1]})}=(\Delta^{higher} f){(v_i^{[2]})}=\\
 &\sum_{\ell_k^{(1,1)}=(v_i,v_j)} f^{[1]}(v_i^{[1]})+f^{[2]}(v_i^{[2]})-f^{[1]}(v_j^{[1]})-f^{[2]}(v_j^{[2]}),\nonumber 
 \eea
The matrix $\mathcal{L}^{higher}$ corresponding to the Laplace operator $\Delta^{higher}$ is given by
\be
\mathcal{L}^{higher}=\left(\begin{array}{cc}
    L^{(1,1)} & L^{(1,1)} \\
    L^{(1,1)} & L^{(1,1)}
\end{array}\right). 
\ee
Hence, the operator $\Delta^{higher}$ and its corresponding matrix $\mathcal{L}^{higher}$ capture the four-body interactions between replica nodes belonging to   multilinks of type $\vec{m}=(1,1)$.
A normalised higher {-order} Laplacian can then be defined as:  
\be
\mathcal{L}^{higher}_{norm}=\left(\begin{array}{cc}
    L^{(1,1)}_{norm} & L^{(1,1)}_{norm} \\
    L^{(1,1)}_{norm} & L^{(1,1)}_{norm}
\end{array}\right). 
\ee
where 
$L^{(1,1)}_{norm}=(D^{(1,1)})^{-\frac{1}{2}}L^{(1,1)}(D^{(1,1)})^{-\frac{1}{2}}$.

The spectrum of  {the higher-order Laplacian}  
$\mathcal{L}^{higher}$ and of 
 {the higher-order normalized Laplacian}  
$\mathcal{L}^{higher}_{norm}$ have the following major properties:
\begin{enumerate}
\item Both $\mathcal{L}^{higher}$ and $\mathcal{L}^{higher}_{norm}$ are semi-positive definite,  
hence their eigenvalues are non-negative.
    \item For all $N$ dimensional vectors $\bm{w}$,  the $2N$-dimensional vector  $\bm{X}=\left( \bm{w},-\bm{w}\right)$ is in the kernel of $\mathcal{L}^{higher}$ and in the kernel of $\mathcal{L}^{higher}_{norm}$ as well.
    \item The operator $\mathcal{L}^{higher}$ admits  {the} eigenvalues ${\lambda}=2\mu$  {with} corresponding eigenvectors $\bm{v}=\left(\bm{u}, \bm{u}\right)$ where  $\mu$ and $\bm{u}$ are eigenvalues and eigenvectors of matrix $L^{(1,1)}$ respectively.
    Similarly,  the operator $\mathcal{L}^{higher}_{norm}$ admits  {the} eigenvalues ${\lambda}=2\mu$  {with} corresponding eigenvectors $\bm{v}=\left(\bm{u}, \bm{u}\right)$ where  $\mu$ and $\bm{u}$ are eigenvalues and eigenvectors of matrix $L^{(1,1)}_{norm}$ respectively.
\item
From the properties (i) and (ii) it follows that if the network formed exclusively by multilinks of type $\vec{m}=(1,1)$ has $p'$ connected components, then  {both} the kernel of $\mathcal{L}^{higher}$ and the kernel of  $\mathcal{L}^{higher}_{norm}$ have 
dimension $N+p'$.
\end{enumerate}

\section{The Hyper-Laplacians of a duplex  network}

Thanks to the lower- and higher {-order} operators introduced in the previous section, we can define the Hyper-Laplace operator of a duplex network. Consider the $0$-cochain $f$ defined on the replica nodes of a duplex network. The Hyper-Laplace operator of the duplex network is then defined as: 
 \begin{equation}
\Delta f\ {(v)}=\Delta^{lower} f\ {(v)}+\delta_{11} \Delta^{higher} f\ {(v)},
\label{dl}
 \end{equation}
  {where the control parameter} $\delta_{11}\geq 0$ 
  {can be used to modulate the strength} 
   {of higher-order diffusion. Notice that,   
  for $\delta_{11}=0$ the Hyper-Laplacian operator reduces  {to} the lower Laplacian operator, while for $\delta_{11}\gg 1$ higher-order diffusion is dominating, and the Hyper-Laplace operator in a first order approximation  {is} given by $\Delta\simeq \delta_{11}\Delta^{higher}$.}
 {The definition of the Hyper-Laplace operator given by Eq.(\ref{dl}) leads to the explicit expression}
 \bea
\hspace*{-20mm} \Delta f\ {(v_i^{[1]})}&=\sum_{\ell_k^{[1]}=(v_i^{[1]},v_j^{[1]})} f^{[1]}(v_i^{[1]})-f^{[1]}(v_j^{[1]})\nonumber \\
\hspace*{-20mm}&+\delta_{11}\sum_{\ell_k^{(1,1)}=(v_i,v_j)} f^{[1]}(v_i^{[1]})+f^{[2]}(v_i^{[2]})-f^{[1]}(v_j^{[1]})-f^{[2]}(v_j^{[2]})\\
\hspace*{-20mm}  \Delta f\ {(v_i^{[2]})}&=\sum_{\ell_k^{[2]}=(v_i^{[2]},v_j^{[2]})} f^{[2]}(v_i^{[2]})-f^{[2]}(v_j^{[2]})\nonumber \\
\hspace*{-20mm}&+\delta_{11}\sum_{\ell_k^{(1,1)}=(v_i,v_j)} f^{[1]}(v_i^{[1]})+f^{[2]}(v_i^{[2]})-f^{[1]}(v_j^{[1]})-f^{[2]}(v_j^{[2]})
 \eea
 Hence, the Laplace operator $\Delta$ captures pairwise interactions between nodes that are connected by a link in a given layer; additionally it embodies four-body interactions between every set of replica nodes connected by  multilinks of type  $(1,1)$.

As a result, the Hyper-Laplacian matrix of a duplex network can be written as
\bea
\mathcal{L}= \mathcal{L}^{lower}+\delta_{11} \mathcal{L}^{higher}=
     \left(\begin{array}{cc}
      L^{[1]}+\delta_{11}L^{(1,1)} &  \delta_{11}L^{(1,1)}\\
     \delta_{11}L^{(1,1)} &  L^{[2]}+\delta_{11}L^{(1,1)}
     \end{array}\right).
\label{eq:hl}
\eea

Here we also define the normalised Hyper-Laplacian $\mathcal{L}^{norm}$ as 
\bea
\mathcal{L}^{norm}&=&D^{-\frac{1}{2}} \mathcal{L}D^{-\frac{1}{2}}\eea
where $D$ indicates the (weighted) hyper-degree matrix defined as follows:
\bea
D=D_1\oplus D_2=\left(\begin{array}{cc}
D_{1} & 0\\
0 &  D_{2}
\end{array}\right),
\eea
with $D_1=D^{(1,0)}+(1+\delta_{11})D^{(1,1)}$ and $D_2=D^{(0,1)}+(1+\delta_{11})D^{(1,1)}$. Therefore, the  diagonal elements  of $D_1$ and $D_2$ are given by the hyper-degrees ${d}^{[1]}_i$ and ${d}^{[2]}_i$ with
\bea
(D_1)_{ii}=d^{[1]}_i=k^{[1]}_i+\delta_{11} k^{(1,1)}_i,\quad
(D_2)_{ii}=d^{[2]}_i=k^{[2]}_i+\delta_{11}k^{(1,1)}_i.
\eea
In other words, the hyper-degree  of a replica node in
 {each of the two layers of} 
a two-layer multiplex network is made up of two contributions: a contribution coming from the  pairwise interactions that the replica node has in its own layer and a contribution coming from the   4-body interactions that the replica nodes has across the two layers thanks to the multilinks of type $(1,1)$.

 The normalised Hyper-Laplacian is defined as 
 \bea
\mathcal{L}^{norm}&=&D^{-\frac{1}{2}} \mathcal{L}D^{-\frac{1}{2}}
\eea
and has explicit expression
\bea
\hspace{-25mm}\mathcal{L}^{norm}&=&\left(\begin{array}{cc} 
D^{-\frac{1}{2}}_1 L^{[1]} D^{-\frac{1}{2}}_1 + \delta_{11}D^{-\frac{1}{2}}_1L^{(1,1)}D^{-\frac{1}{2}}_1 & \delta_{11}D^{-\frac{1}{2}}_1L^{(1,1)}D^{-\frac{1}{2}}_2\\
\delta_{11}D^{-\frac{1}{2}}_2 L^{(1,1)}D^{-\frac{1}{2}}_1 & D^{-\frac{1}{2}}_2 L^{[2]} D^{-\frac{1}{2}}_2 + \delta_{11}D^{-\frac{1}{2}}_2 L^{(1,1)}D^{-\frac{1}{2}}_2\end{array}\right),
\eea
where we adopt the convention that if the hyper-degree of a replica node is zero (i.e. $d_i^{[\alpha]}=0$), then $(D_{\alpha}^{-1})_{ii}=0$.
While the normalised Laplacian $\mathcal{L}^{norm}$ is symmetric, it is also possible to consider the isospectral asymmetric normalised Laplacian $\mathcal{L}^{asym}$ defined as 
\bea
\mathcal{L}^{asym}=D^{-1}\mathcal{L}
\eea
that has the explicit expression 
\bea
\hspace{-25mm}\mathcal{L}^{asym}&=&\left(\begin{array}{cc} 
D^{-1}_1 L^{[1]}  + \delta_{11}D^{-1}_1L^{(1,1)} & \delta_{11}D^{-1}_1L^{(1,1)}\\
\delta_{11}D^{-1}_2 L^{(1,1)} & D^{-1}_2 L^{[2]}  + \delta_{11}D^{-1}_2 L^{(1,1)}
\end{array}\right).
\eea

\begin{figure}[!htbp]
\begin{center}
$\hspace*{10mm}\includegraphics[width=0.9\textwidth]{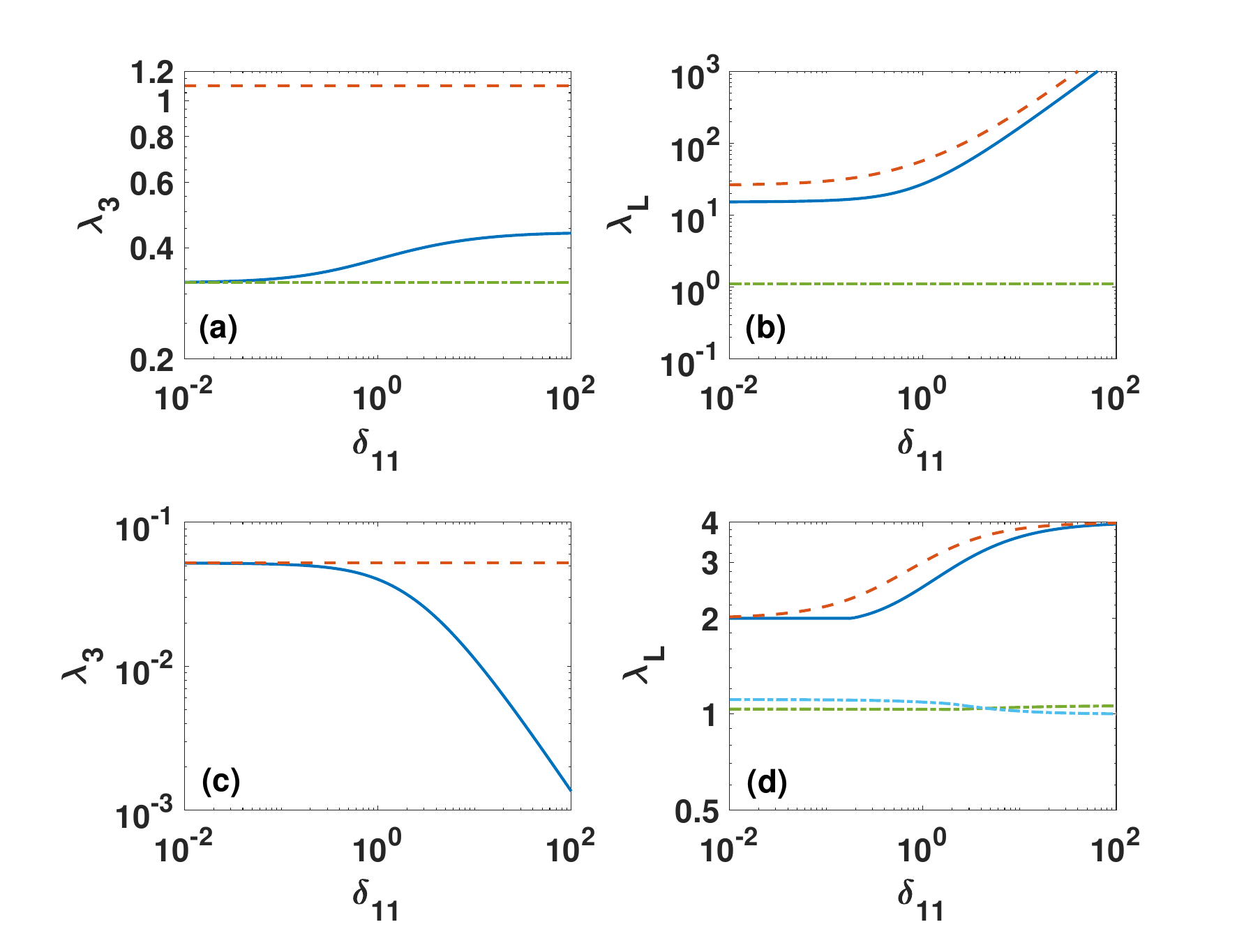}
$
\caption{{\bf Bounds for the Fielder and the largest eigenvalue of the Hyper-Laplacians.} The Fiedler eigenvalue $\lambda_3$ and the largest eigenvalue $\lambda_L$ (solid blue lines) are plotted versus the weight of higher-order interactions for the un-normalised Hyper-Laplacian (panels (a) and (b)) and for the normalised Hyper-Laplacian (panels (c) and (d)). The red dashed lines indicate the upper bounds, while the green and cyan dot-dashed lines indicate the lower bounds. The duplex networks considered have  Poisson multidegree distributions with averages $\Avg{k^{(1,0)}}=2, \Avg{k^{(0,1)}}=4, \Avg{k^{(1,1)}}=2$ (panels (a) and (b)) and $\Avg{k^{(1,0)}}=2, \Avg{k^{(0,1)}}=2.8, \Avg{k^{(1,1)}}=0.7$ (panels (c) and (d)). The upper and lower bounds in panel (a) are given by Eq. (\ref{ineq:lambda3}). The upper bounds in panels (b) and (d) are given by Eq.~(\ref{lambdaL}) and Eq.~(\ref{lambdaL2}). The lower bounds in panels (b) and (d) are given by $1+{1}/{d_{max}}$ and by $\max\left(\frac{\Avg{d^{[1]}}N}{\Avg{d^{[1]}}N-d^{[1]}_{max}} , \frac{\Avg{d^{[2]}}N}{\Avg{d^{[2]}}N-d^{[2]}_{max}}\right)$. The upper bound in panel (b) is given by Eq. (\ref{lambda_3_norm_upper}).}
\label{fig:2}
\end{center}
\end{figure}
\section{Spectral properties of Hyper-Laplacians}
\label{properties}
In this section we investigate the  {basic} spectral properties of the Hyper-Laplacians defined in the previous section, and we derive upper and lower bounds for the smallest non zero eigenvalue and for the largest eigenvalue using the  Rayleigh quotient \cite{chung1997spectral}.
Let us first start with investigating the spectral properties of  {the Hyper-Laplacian matrix}  $\mathcal{L}$  {defined in Eq.~(\ref{eq:hl})}. 
Given a generic column vector $\bm{X}=\left(
 \bm{x}^{[1]},\bm{x}^{[2]}
\right)^{\top}$ defined on the replica nodes of a duplex network, the Raylegh quotient is defined as
\bea
RQ(\mathcal{L})=\frac{\bm{X}^T\mathcal{L}\bm{X}}{\bm{X}^T\bm{X}},
\eea
where $\bm{X}^{\top}\mathcal{L}\bm{X}$ is given by
\bea
\bm{X}^{\top}\mathcal{L}\bm{X}&=&\frac{1}{2}\left[\sum_{i,j}a^{[1]}_{ij}(x^{[1]}_i-x^{[1]}_j)^2
     +\sum_{i,j}a^{[2]}_{ij}(x^{[2]}_i-x^{[2]}_j)^2\right]\nonumber \\
     && + \frac{1}{2}\delta_{11}\left[\sum_{i,j}A^{(1,1)}_{ij}(x^{[1]}_i-x^{[1]}_j+x^{[2]}_i-x^{[2]}_j)^2\right].
\eea
As it is evident from the Rayleigh quotient, since $\bm{X}^{\top}\mathcal{L}\bm{X}\geq 0$, the Hyper-Laplacian is a positive semi-definite matrix and consequently its eigenvalues are non-negative. 
From the investigation  of the Rayleigh quotient we can deduce the following  properties of the Hyper-Laplacian:
\begin{enumerate}
    \item The Hyper-Laplacian $\mathcal{L}$ always admits a zero eigenvalue $\lambda=0$ whose algebraic multiplicity is equal  {to} the sum of the  connected components of the two layers (see \ref{Ap:kernel} for the proof).
    \item  The smallest non-zero eigenvalue $\lambda_s(\mathcal{L})$ and $\lambda_L$ and  the largest eigenvalue of $\lambda_L(\mathcal{L})$, satisfy (see \ref{Ap:gen} for the proof)
  \begin{equation}
 \lambda_s(\mathcal{L}) \leq \frac{\sum_{i=1}^N(d_i^{[1]}+d_i^{[2]})}{2N-(s-1)}\leq \lambda_L(\mathcal{L}),
 \label{gen1}
  \end{equation}
  where $s-1$ indicates the degeneracy of the zero eigenvalue.
  \item If each layer is formed by a  {single connected component}, the degeneracy of the   eigenvalue $\lambda=0$ is two. The smallest non-zero eigenvalue (Fiedler eigenvalue) of non-normalised Hyper-Laplacian $\lambda_3(\mathcal{L})$ has the following bounds (see \ref{Ap:l3} for the proof):
 {
 \bea
    \lambda_3(\mathcal{L}^{lower})\leq\lambda_s(\mathcal{L})\leq \frac{1}{2}
\lambda_2\left({L^{[1]}+L^{[2]}}\right),
    \label{ineq:lambda3}
 \eea
where  $\lambda_2\left(L^{[1]}+L^{[2]}\right)$ is the Fiedler eigenvalue of the aggregated network, while $ \lambda_3(\mathcal{L}^{lower})$ is the Fiedler eigenvalue of $\mathcal{L}^{lower}$ given by Eq. (\ref{lower_l2}).}
\item
The largest eigenvalue $\lambda_L$ of the non-normalized Hyper-Laplacian admits as upper bound (see \ref{Ap:lambdaL} for the proof)
\begin{equation}
    \lambda_L(\mathcal{L}) \leq \left(2+\frac{2\delta_{11}}{1+\delta_{11}}\right) d_{max},
    \label{lambdaL}
\end{equation}
where $d_{max}$ is the largest  hyper-degree  of the multiplex network, i.e.
\bea
d_{max}=\max\left(\max_{i=1,\dots,N}d_i^{[1]},\max_{i=1,\ldots,N} d_i^{[2]}\right).
\eea
Note that in Eq. (\ref{lambdaL}) the equality holds if and only if two layers are identical and bipartite.
\end{enumerate}

 {A Rayleigh quotient can also be used}  to investigate the spectrum of the normalised  Hyper-Laplacian $\mathcal{L}^{norm}$ \cite{chung1997spectral}.  {Again,} 
given a generic  column vector $\bm{X}$ defined on the replica nodes of the considered duplex network, the Rayleigh quotient of the normalised Hyper-Laplacian is: 
 \bea\label{raylnorm0}
     &\hspace{-20mm} RQ(\mathcal{L}^{norm})=\frac{\bm{X}^T\mathcal{L}^{norm}\bm{X}}{\bm{X}^T\bm{X}}=\frac{\bm{X}^TD^{-\frac{1}{2}} \mathcal{L} D^{-\frac{1}{2}} \bm{X}}{\bm{X}^T \bm{X}}=\frac{\bm{Y}^T \mathcal{L} \bm{Y}}{\bm{Y}^T D \bm{Y}}
      \eea
 where $\bm{Y}=D^{-\frac{1}{2}}\bm{X}$.
 As evident from the Rayleigh quotient, the normalised Hyper-Laplacian is a positive semi-definite matrix and consequently its eigenvalues are non-negative. 
 Moreover,  the normalised Hyper-Laplacian has the following properties:

\begin{enumerate}
\item The normalised Hyper-Laplacian $\mathcal{L}^{norm}$ always admits a zero eigenvalue $\lambda=0$ whose algebraic multiplicity is equal  {to} the sum of  {the number of} connected components of two layers (see \ref{Ap:kernel} for the proof). 
  \item  The smallest non-zero eigenvalue $\lambda_s(\mathcal{L}^{norm})$ and $\lambda_L$ and  the largest eigenvalue of $\lambda_L(\mathcal{L}^{norm})$, satisfy (see \ref{Ap:gen} for the proof)
  \begin{equation}
 \lambda_s(\mathcal{L}^{norm}) \leq \frac{N}{N-(s-1)/2}\leq \lambda_L(\mathcal{L}^{norm}),
 \label{gen2}
  \end{equation}
  where $s-1$ indicates the degeneracy of the zero eigenvalue.
  \item  Let assume that each layer of the duplex network is connected. Without loss of generality,  we assume that the first layer has smaller  Fiedler eigenvalue of the second layer, i.e. 
  \bea
  \lambda_2(L^{[1]}_{norm})\leq \lambda_2(L^{[2]}_{norm}).
  \eea
  As long as 
   \bea
  \frac{\bm{u}_1^{\top}L^{(1,1)}\bm{u}_1}{\bm{u}_1^{\top}D^{(1,1)}\bm{u}_1}\leq \frac{\bm{u}_1^{\top}L^{(1,0)}\bm{u}_1}{\bm{u}_1^{\top}D^{(1,0)}\bm{u}_1},\eea { or equivalently } 
  \bea
  \frac{\bm{u}_1^{\top}L^{(1,1)}\bm{u}_1}{\bm{u}_1^{\top}D^{(1,1)}\bm{u}_1}\leq  \lambda_2(L^{[1]}_{norm}),
  \eea
  where $\bm{u}_1=[D^{[1]}]^{-\frac{1}{2}}\bm{y}_1$  with $\bm{y}_1$ indicating the eigenvector corresponding to the Fiedler eigenvalue of $L^{[1]}_{norm}$, then we  have (see \ref{Ap:upper2} for the proof) 
  \bea
   \lambda_3(\mathcal{L}^{norm})\leq  \lambda_2(L^{[1]}_{norm}).
   \label{lambda_3_norm_upper}
  \eea 
  \item  In addition to the bound listed in (i), by using  similar approach as in  \cite{li2014bounds,grone1994laplacian}, we can prove (see \ref{Ap:lambdaL2}) that the largest eigenvalue of the normalised Hyper-Laplacian $\mathcal{L}^{norm}$ has the following upper bound
\begin{equation}
    \lambda_L(\mathcal{L}^{norm})   \geq \max{ \Bigg\{\frac{\Avg{d^{[1]}}N}{\Avg{d^{[1]}}N-d^{[1]}_{max}} , \frac{\Avg{d^{[2]}}N}{\Avg{d^{[2]}}N-d^{[2]}_{max}},1+\frac{1}{d_{max}} \Bigg\}},
    \label{lambdaL_norm_upper}    
\end{equation}
where $d_{max}$ is the largest  hyper-degree in two layers of multiplex network.
\item The largest eigenvalue of the normalised Hyper-Laplacian $\lambda_L(\mathcal{L}^{norm})$ admits as upper bound (see \ref{Ap:lambdaL} for the proof)
\begin{equation}
    \lambda_L(\mathcal{L}^{norm}) \leq 2+\frac{2\delta_{11}}{1+\delta_{11}},
    \label{lambdaL2}
\end{equation}
with equality if and only if two layers are identical and bipartite.
\end{enumerate}

Finally,  the asymmetric normalised Laplacian $\mathcal{L}^{asym}$ can be easily shown to have  the same   spectrum as $\mathcal{L}^{norm}$. Additionally  the  left eigenvector of $\mathcal{L}^{asym}$  associated to the eigenvalue $\lambda$  is given by $\bm{v}^{asym,L}=D^{-\frac{1}{2}}\bm{v}^{norm}$ where $\bm{v}^{norm}$ is the eigenvector of $\mathcal{L}^{norm}$ associated to the same eigenvalue, while the right eigenvalue associated to the eigenvalue $\lambda$ is given by $\bm{v}^{asym,R}=D^{\frac{1}{2}}\bm{v}^{norm}$.
Note that, when both layers are connected, the asymmetric normalised Laplacian has a twice degenerate eigenvalue $\lambda=0$ with associated left eigenvectors taking constant values on the replica nodes of layer $1$ and layer $2$ respectively, while the right eigenvectors have elements proportional to the hyper-degrees of layer $1$ and the hyper-degrees of layer $2$ respectively. 

In Figure $3$,  we show the Fielder eigenvalue and the largest eigenvalue of both the un-normalised Hyper-Laplacian $\mathcal{L}$ and the normalised Hyper-Laplacian $\mathcal{L}^{norm}$ as a function of $\delta_{11}$. The considered duplex networks are networks with Poisson distribution of multilinks \cite{bianconi2013statistical,bianconi2018multilayer,Git_repository}.
In addition to the dependence of the eigenvalues of the Hyper-Laplacian  {on} the diffusion constant $\delta_{11}$, the figure also shows the upper and lower bound theoretically predicted in this section showing that for the duplex networks considered in the figure, the upper bounds (Eq. (\ref{lambdaL}) and Eq. (\ref{lambdaL2})) on the largest eigenvalues are tight, and that also the upper bound  (Eq.(\ref{lambda_3_norm_upper})) on $\lambda_3(\mathcal{L}^{norm})$ is tight for $\delta_{11}\ll 1$.

\begin{figure}[!htbp]
\begin{center}
\includegraphics[width=0.9\textwidth]{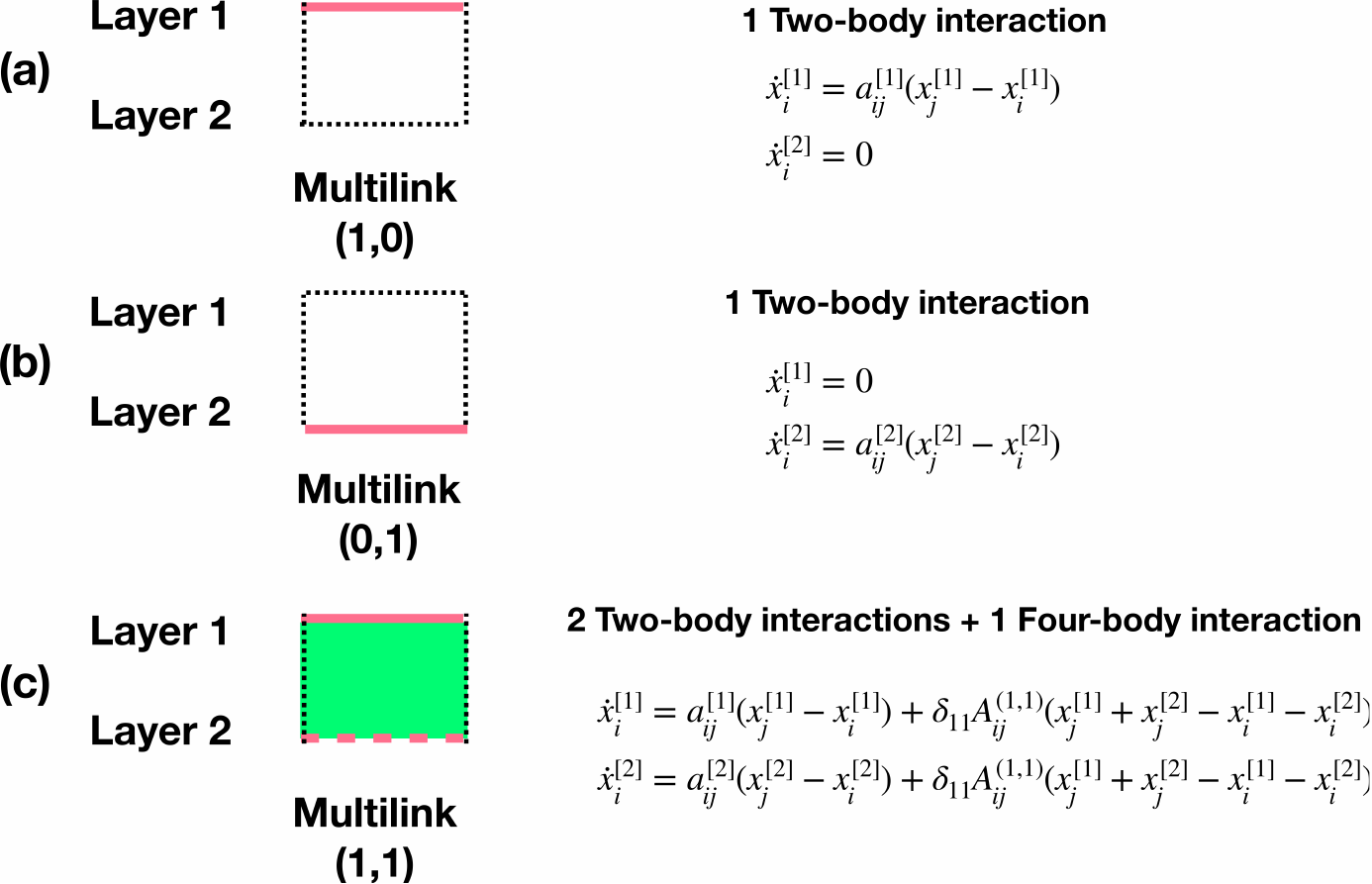}
\caption{{\bf Schematic representation of the hyper-diffusion on a multiplex network.} Any two nodes connected by a multilink of type $(1,0)$ (panel (a)) or of type  $(0,1)$ (panel (b)) are coupled by the lower Laplacian only encoding for the pairwise interactions between replica nodes. However any two nodes connected by a multilink of type $(1,1)$ (panel (c)) are coupled by the higher Laplacian as well, encoding for four-body interactions and coupling the dynamics among the two layers.  }
\label{fig:3}
\end{center}
\end{figure}

\section{Hyper-diffusion on a duplex network}

Hyper-diffusion  refers to a diffusion process driven by the Hyper-Laplacians $\mathcal{L}$ and $\mathcal{L}^{asym}$ and capturing both intra-layer (pairwise) interactions and  inter-layer 4-body interactions encoded by the multilinks of type $(1,1)$.
Let us start by discussing the hyper-diffusion driven by the non-normalised Hyper-Laplacian $\mathcal{L}$.
We assume that at time $t$ the state of the replica nodes is characterized by the $2N$-dimensional vector $\bm{X}(t)=\left(\bm{x}^{[1]}(t),\bm{x}^{[2]}(t)\right)$, 
 {and that the system is initially in the state 
$\bm{X}(0)=\bm{X}_0=\left(\bm{x}^{[1]}_0,\bm{x}^{[2]}_0\right)$.}
 {The evolution of hyper-diffusion is governed by the following differential equation:} 
%
  \begin{equation}\label{diffusiton}
  \frac{d\bm{X}}{dt}=-\mathcal{L}\bm{X}=\left(\mathcal{L}^{lower}+{\delta_{11}}\mathcal{L}^{higher}\right)\bm{X},
  \end{equation}
which element-wise reads 
 \bea
\hspace*{-15mm}\frac{d{x}^{[1]}_i}{dt}=\sum_{j=1}^{N} a^{[1]}_{ij}(x^{[1]}_j-x^{[1]}_i)+{\delta_{11}}\sum_{j=1}^{N} A^{(1,1)}_{ij}(x^{[1]}_j+x^{[2]}_j-x^{[1]}_i-x^{[2]}_i)
\nonumber
\\
\hspace*{-15mm}\frac{d{x}^{[2]}_i}{dt}=\sum_{j=1}^{N} a^{[2]}_{ij}(x^{[2]}_j-x^{[2]}_i)+{\delta_{11}}\sum_{j=1}^{N} A^{(1,1)}_{ij}(x^{[1]}_j+x^{[2]}_j-x^{[1]}_i-x^{[2]}_i)
\label{dif2}
 \eea
  {with $i=1,2,\ldots,N$.}  
 For $\delta_{11}=0$ the diffusion in the two layers is uncoupled, while for $\delta_{11}>0$ the presence of multilinks of type $(1,1)$ embodying higher-order interactions between replica nodes allows hyper-diffusion to couple the dynamics on the two layers as sketched  
 in Figure $\ref{fig:3}$.
 Note that the average state $ \Avg{x^{[\alpha]}(t)}$ of the replica nodes of each layer  $\alpha$, defined as
 \bea
 \Avg{x^{[\alpha]}(t)}=\frac{1}{N}\sum_{i=1}^Nx_i^{[\alpha]}(t)
 \eea
 is a conserved property of the dynamics, as the dynamical Eqs.(\ref{dif2}) implies
 \bea
 \frac{d\Avg{x^{[\alpha]}(t)}}{dt}=0.
 \eea
 Therefore,  we have that  {the average state $ \Avg{x^{[\alpha]}(t)}$ of the replica nodes in layer $\alpha$ is  {constant in} 
 time, i.e. }
 \bea
 \Avg{x^{[\alpha]}(t)}=\Avg{x^{[\alpha]}_0} \ \forall t\geq 0.
 \eea
 Given initial condition $\bm{X}(t=0)=\bm{X}_0$, the dynamical system defined in Eq. (\ref{diffusiton})  has solution
\begin{equation}
     \bm{X}(t)=e^{-\mathcal{L}t}\bm{X}_0.
\end{equation}
If we decompose initial condition $\bm{X}_0=(\bm{x}^{[1]}_0,\bm{x}^{[2]}_0)$ into the basis of eigenvectors $\bm{v}_n$ of Hyper-Laplacian, i.e. $\bm{X}_{0}=\sum_n c_n\bm{v}_n$, the {solution} to Eq.~(\ref{diffusiton}) reads 
\begin{equation}
    \bm{X}(t)=\sum_{n} e^{- \lambda_n t}c_{n} \bm{v}_n.
\end{equation}
Assuming that two layers are connected, as long as the spectrum of the  {Hyper-Laplacian} displays a spectral gap, the {solution} to Eq.~(\ref{diffusiton}) reads 
can be approximated as
\begin{equation}
    \bm{X}(t)\simeq c_1\bm{v}_1+c_2\bm{v}_2+c_3e^{-\lambda_3 t}\bm{v}_3,
\end{equation}
where $\bm{v}_1=({\bf 1}/\sqrt{N},\bm{0})^{\top}$ and $\bm{v}_2=(\bm{0},{\bf 1}/\sqrt{N})^{\top}$ are the  eigenvectors of Hyper-Laplacian corresponding to eigenvalue $\lambda=0$, which are homogeneous on the replica nodes of the first and second layer respectively, and  $\bm{v}_3$ is the eigenvector associated to the first non-zero eigenvalue of Hyper-Laplacian $\lambda_3(\mathcal{L})$. 
Hence, in the presence of a finite spectral gap of $\mathcal{L}$, the  relaxation time scale $\tau$ of  hyper-diffusion  is given by 
\begin{equation}
\tau=\frac{1}{\lambda_3(\mathcal{L})},
\end{equation}
and the steady state is given by: 
\bea
\lim_{t\to \infty}{\bm{X}}=\left(\Avg{x^{[1]}_0}{\bf 1},\Avg{x^{[2]}_0}{\bf 1}\right)^{\top},
\eea
where $\Avg{x^{[\alpha]}_0}$ indicates the average of the elements of $\bm{x}_0^{[\alpha]}$.
Therefore, as long as the initial states of the replica nodes of layer $1$ is drawn from the same distribution of the initial state of the replica nodes in layer $2$, in the limit of $N\to\infty$ the final state on all the replica nodes of the duplex network  {will be} 
the same.
 
The hyper-diffusion driven by asymmetric normalised Hyper-Laplacian $\mathcal{L}^{asym}$ can be defined in an analogous way. Again, given the  initial condition  $\bm{X}(0)=\bm{X}_0=\left(\bm{x}^{[1]}_0,\bm{x}^{[2]}_0\right)$, 
the dynamics obeys the system of equations  
  \begin{equation}\label{diffusionb}
  \frac{d\bm{X}}{dt}=-\mathcal{L}^{asym}\bm{X},
  \end{equation}
which element-wise reads 
 \bea
\hspace*{-15mm}\frac{d{x}^{[1]}_i}{dt}=\frac{1}{d_i^{[1]}}\sum_{j=1}^{N} a^{[1]}_{ij}(x^{[1]}_j-x^{[1]}_i)+\delta_{11}\frac{1}{d_i^{[1]}}\sum_{j=1}^{N} A^{(1,1)}_{ij}(x^{[1]}_j+x^{[2]}_j-x^{[1]}_i-x^{[2]}_i)
\nonumber
\\
\hspace*{-15mm}\frac{d{x}^{[2]}_i}{dt}=\frac{1}{d_i^{[2]}}\sum_{j=1}^{N} a^{[2]}_{ij}(x^{[2]}_j-x^{[2]}_i)+{\delta_{11}}\frac{1}{d_i^{[2]}}\sum_{j=1}^{N} A^{(1,1)}_{ij}(x^{[1]}_j+x^{[2]}_j-x^{[1]}_i-x^{[2]}_i)
\label{dif2b}
 \eea
 {with $i=1,2,\ldots,N$.}  
 Now it is immediate to show that the conserved quantities are 
 \bea
 \Avg{x^{[\alpha]}(t)}_{d^{[\alpha]}}=\frac{1}{\Avg{d^{[\alpha]}}N}\sum_{i=1}^Nd_i^{[\alpha]}x_i^{[\alpha]}(t)
 \eea
 as the dynamical Eqs.(\ref{dif2b}) implies
 \bea
 \frac{d\Avg{x^{[\alpha]}(t)}_{d^{[\alpha]}}}{dt}=0.
 \eea
 Therefore we have that 
 \bea
 \Avg{x^{[\alpha]}(t)}_{d^{[\alpha]}}=\Avg{x^{[\alpha]}_0}_{d^{[\alpha]}} \ \forall t\geq 0.
 \eea
 Given the initial condition $\bm{X}(t=0)=\bm{X}_0$, the dynamical system defined in Eq. (\ref{diffusiton})  has solution
\begin{equation}
     \bm{X}(t)=e^{-\mathcal{L}^{asym}t}\bm{X}_0.
\end{equation}
If we decompose initial condition $\bm{X}_0=(\bm{x}^{[1]}_0,\bm{x}^{[2]}_0)$ into the basis of left eigenvectors $\bm{v}_n^{L}$ of the asymmetric normalised Hyper-Laplacian, i.e. $\bm{X}_{0}=\sum_n c_n\bm{v}_n^{L}$, Eq. (\ref{diffusiton}) reads
\begin{equation}
    \bm{X}(t)=\sum_{n} e^{- \lambda_n t}c_{n} \bm{v}_n^{L}.
\end{equation}
Therefore, assuming that two layers are connected, as long as the spectrum of the Hyper-Laplacian displays a spectral gap, the solution to the equation Eq. (\ref{diffusiton}) can be approximated as:
\begin{equation}
    \bm{X}(t)\simeq c_1\bm{v}_1^L+c_2\bm{v}_2^L+c_3e^{-\lambda_3 t}\bm{v}_3^L,
\end{equation}
where $\bm{v}_1^L=({\bf 1},\bm{0})^{\top}$ and $\bm{v}_2^L=(\bm{0},{\bf 1})^{\top}$ are the  left eigenvectors of Hyper-Laplacian corresponding to eigenvalue $\lambda=0$, which are homogeneous on the replica nodes of the first and second layer respectively, $c^{[\alpha]}=\Avg{x^{[\alpha]}_0}_{d^{[\alpha]}}$ and  $\bm{v}_3^L$ is the left eigenvector associated to the first non-zero eigenvalue of Hyper-Laplacian $\lambda_3(\mathcal{L})$. 

Hence, in the presence of a finite spectral gap of $\mathcal{L}$, the  relaxation time scale $\tau$ of  hyper-diffusion  is given by 
\begin{equation}
\tau=\frac{1}{\lambda_3(\mathcal{L}^{asym})},
\end{equation}
and the steady state is 
\bea
\lim_{t\to \infty}{\bm{X}}=\left(\Avg{x^{[1]}_0}_{d^{[1]}}{\bf 1},\Avg{x^{[2]}_0}_{d^{[2]}}{\bf 1}\right)^{\top}.
\eea

\begin{figure}[!htbp]
\begin{center}
$\hspace*{15mm}\includegraphics[width=0.9\textwidth]{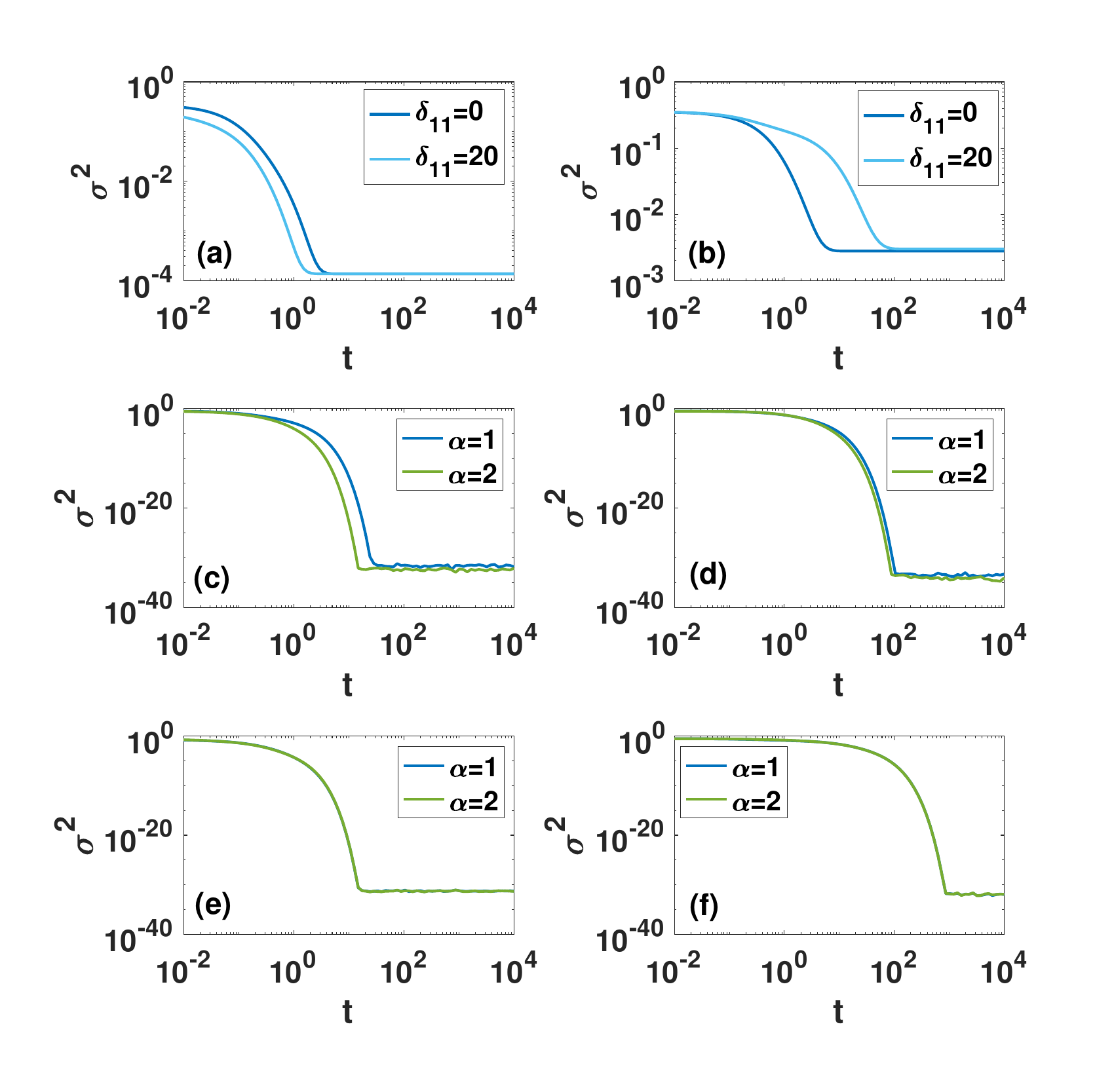}
$
\caption{{\bf Convergence of hyper-diffusion to the steady state.} The variance $\sigma^2$ 
 {of} the states of the replica nodes
is plotted  {as a function of time} 
in absence ($\delta_{11}=0$) and in presence 
($\delta_{11}=20$) of coupling between the layers 
for hyper-diffusion driven by the un-normalised  Hyper-Laplacian $\mathcal{L}$ (panel a) and for the normalised asymmetric Hyper-Laplacian $\mathcal{L}^{asym}$ (panel (b)). Note that the relaxation to the steady state for $\delta_{11}=0$ is driven by $\mathcal{L}^{lower}$ in panel (a) and by $D^{-\frac{1}{2}}\mathcal{L}^{lower}D^{-\frac{1}{2}}$ in panel (b), that are matrices with different Fielder eigenvalues, hence the two relaxation curves are different.
The variances $\sigma^2$ of the states of the replica nodes of layer $\alpha$, with $\alpha\in \{1,2\}$, are plotted in absence of hyper-diffusion ($\delta_{11}=0$) (panel (c) and (d)) in presence of hyper-diffusion ($\delta_{11}=20$) (panels (e) and (f)) for the un-normalised Hyper-Laplacian $\mathcal{L}$ (panels (c) and (e)) and for the normalised asymmetric Hyper-Laplacian $\mathcal{L}^{asym}$ (panels (d) and (f)). In panels (e) and (f) the two curves are super-imposed.
The duplex networks have $N=100$ nodes and Poisson distribution of multilinks with average
$\Avg{k^{(1,0)}}=2, \Avg{k^{(0,1)}}=4, \Avg{k^{(1,1)}}=2$ (panel (a));
$\Avg{k^{(1,0)}}=2, \Avg{k^{(0,1)}}=2, \Avg{k^{(1,1)}}=5$ (panel (b));
$\Avg{k^{(1,0)}}=3, \Avg{k^{(0,1)}}=5, \Avg{k^{(1,1)}}=3$ (panels (c)-(f).
}
\label{fig:4}
\end{center}
\end{figure}
Independent of the choice of the Hyper-Laplacian, we observe the following  important property of hyper-diffusion: although the dynamical state of each layer has an average (or a weighted average) that is preserved, the dynamics of the two layers is non-trivially coupled.  {In particular  the coupling between the layers is revealed by the relaxation time to the steady state of the dynamics that becomes synchronous for sufficiently large values of $\delta_{11}$.}
In fact, for $\delta_{11}=0$ the layers are uncoupled, and when the topology of the two layers is different, the characteristic relaxation to the steady state  is different with each layer $\alpha$.  {Indeed each layer $\alpha$ has a characteristic relaxation time given by the inverse of the Fiedler eigenvalue of $L^{[\alpha]}$.}
However when $\delta_{11}>0$,  the characteristic relaxation time  of the two layers becomes the same and is given by the inverse of the Fiedler eigenvalue of the Hyper-Laplacian as long as the Fiedler eigenvector is non-localised on a single layer  {(which typically occurs for sufficiently large values of $\delta_{11}$)}.
This phenomenon is very evident if we monitor the variance $\sigma^2$ of the state of the replica nodes in each single layer as a function of $t$ in the two limiting cases where $\delta_{11}=0$ and $\delta_{11}>0$ (see Figure $\ref{fig:4}$ for an example).
However ,{} we notice an important difference between hyper-diffusion driven by the un-normalised Hyper-Laplacian and the one driven by the asymmetric normalised Hyper-Laplacian.   The Fiedler eigenvalue of $\mathcal{L}$ increases with $\delta_{11}$ implying that hyper-diffusion ($\delta_{11}>0$) driven by $\mathcal{L}$  is faster than diffusion of the slowest of the two layers and becomes increasingly faster by increasing $\delta_{11}$ (see Figure $\ref{fig:4}$). For some duplex network topologies hyper-diffusion driven by the un-normalised Hyper-Laplacian $\mathcal{L}$ can also lead to super-diffusion, i.e. diffusion faster than the diffusion of the fastest layer of the duplex network (see Figure $\ref{fig:4}(c)$ for $\delta_{11}=0$ and Figure $\ref{fig:4}(e)$ for $\delta_{11}=20$).
However,  the  Fiedler eigenvalue of $\mathcal{L}^{norm}$ decreases with $\delta_{11}$ as a consequence of the fact that the hyperdegree also depends on $\delta_{11}$. Therefore hyper-diffusion driven by the normalised Hyper-Laplacian becomes slower as $\delta_{11}$ is increased (see Figure $\ref{fig:4}$).

\section{Asymptotic behaviour of hyper-diffusion}
The properties of hyper-diffusion are dictated by the spectrum of the Hyper-Laplacians. 
In this section we investigate the asymptotic behaviour  {of} hyper-diffusion driven by the Hyper-Laplacians  in the limiting cases $\delta_{11}\ll1$ and $\delta_{11}\gg1$. To this end,  we perform a perturbative expansion \cite{cohen1986quantum} of the  eigenvalues of the Hyper-Laplacians under the assumption that each layer is formed by a connected network.
\subsection{Asymptotic expansion for $\delta_{11}\ll 1$}
Let us consider first the asymptotic expansion of the eigenvalues of the un-normalised Hyper-Laplacian $\mathcal{L}$ and subsequently generalise the results obtained to the normalised Hyper-Laplacian $\mathcal{L}^{norm}$, which has the same spectrum of $\mathcal{L}^{asym}$ as observed in Sec. \ref{properties}.

For  {$\delta_{11}$ small}, the non-zero eigenvalues of Hyper-Laplacian $\mathcal{L}=\mathcal{L}^{lower}+\delta_{11} \mathcal{L}^{higher}$ are perturbations of non-zero eigenvalues of $\mathcal{L}^{lower}$. Let $\bm{x}^{[\alpha]}$ be a  normalised eigenvector of Laplacian of layer $\alpha$ with non-degenerate eigenvalue $\lambda^{[\alpha]}$. The eigenvector $\bm{x}^{[\alpha]}$ satisfies,
\begin{equation}
    L^{[\alpha]}\bm{x}^{[\alpha]}=\lambda^{[\alpha]}\bm{x}^{[\alpha]}.
\end{equation}
The corresponding eigenvector of $\mathcal{L}^{lower}$ is $\bm{v}^{[\alpha]}=\bm{e}_{\alpha}\otimes \bm{x}^{[\alpha]}$, where $\bm{e}_{\alpha}$ is a $2$-dimensional canonical vector whose $\alpha$th component is one. Hence $\bm{v}^{[\alpha]}$ satisf {ies}
\begin{equation}
    \mathcal{L}^{lower} \bm{v}^{[\alpha]}=\lambda^{[\alpha]} \bm{v}^{[\alpha]}.
    \label{eig_alpha}
\end{equation}
For $\delta_{11}\ll1$, using a first-order perturbative expansion of the eigenvector $\bm{v}$ of $\mathcal{L}$ and its corresponding eigenvalue $\lambda$  we obtain 
\bea
&\bm{v}\approx\bm{v}^{[\alpha]}+\delta_{11}\bm{v}',\\
&\lambda\approx\lambda^{[\alpha]}+\delta_{11}\lambda',
\label{per_ll}
\eea
where in order to guarantee that $\bm{v}$ has norm equal to one in the first order of the perturbative expansion, we must impose that $\bm{v}'$ is orthogonal to $\bm{v}^{[\alpha]}$.
Substituting $\bm{v}$ and $\lambda$ in  $\mathcal{L}\bm{v}=\lambda\bm{v}$ yields in the first order approximation
\begin{equation}\label{eig}
    \mathcal{L}^{lower}\bm{v}'+ \mathcal{L}^{higher}\bm{v}^{[\alpha]}=\lambda^{[\alpha]}\bm{v}'+\lambda'\bm{v}^{[\alpha]}.
\end{equation}
By multiplying both sides of Eq. (\ref{eig}) from the left by $(\bm{v}^{[\alpha]})^{\top}$, using $(\bm{v}^{[\alpha]})^{\top}\mathcal{L}^{lower}=\lambda^{[\alpha]}(\bm{v}^{[\alpha]})^{\top}$, we can derive the following expression for $\lambda'$
\bea
    \lambda'=\frac{\bm{x}^{[\alpha]\top} L^{(1,1)} \bm{x}^{[\alpha]}}{\bm{x}^{[\alpha],\top} \bm{x}^{[\alpha]}}=RQ(L^{(1,1)},\bm{x}^{[\alpha]}).
\eea
Hence, for  $\delta_{11}\ll 1$, the non-zero eigenvalues of Hyper-Laplacian are given by
\begin{equation}\label{nonzero_small}
    \lambda \approx \lambda^{[\alpha]}+RQ(L^{(1,1)},\bm{x}^{[\alpha]})\delta_{11} ,
\end{equation}
where $\lambda^{[\alpha]}$ and $\bm{x}^{[\alpha]}$ are the eigenvalues and respective eigenvectors of Laplacian of layer $\alpha$.
Following a similar procedure we can generalise the results to the normalized Hyper-Laplacian  $\mathcal{L}^{norm}$ finding that the eigenvalues $\lambda'$ of $\mathcal{L}^{norm}$, in the first order of perturbation theory, can be expressed as 
\bea
    \lambda=\lambda^{[\alpha]}+RQ(D_{\alpha}^{-\frac{1}{2}}L^{(1,1)}D_{\alpha}^{-\frac{1}{2}},\bm{x}^{[\alpha]}),
    \label{norm_ndll1}
\eea
where here $\bm{x}^{[\alpha]}$ indicates the eigenvalue of the normalised Laplacian $L^{[\alpha]}_{norm}$ and $\lambda^{[\alpha]}$ indicates its corresponding (non degenerate) eigenvalue.

Note that Eq.(\ref{nonzero_small}) and Eq.(\ref{norm_ndll1}) are only valid if the eigenvalue $\lambda^{[\alpha]}$ is non degenerate. For the asymptotic expansion in case in which $\lambda^{[\alpha]}$ is degenerate, see \ref{Ap:degenerate}.
\subsection{Asymptotic expansion for $\delta_{11}\gg1$}
Let us now consider the asymptotic expansion of the spectrum of the Hyper-Laplacians for $\delta_{11}\gg 1$.
Let us first discuss this expansion for the un-normalised Hyper-Laplacian $\mathcal{L}$ and then discuss the differences with the asymptotic expansion of the spectrum of the normalised Hyper-Laplacian $\mathcal{L}^{norm}$.

For $\delta_{11}\gg 1$, the Hyper-Laplacian $\mathcal{L}$ can be written as
\begin{equation}
    \mathcal{L}=\delta_{11}(\mathcal{L}^{higher}+\epsilon \mathcal{L}^{lower})=\delta_{11}\hat{{\mathcal{L}}},
\end{equation}
where $\epsilon=1/\delta_{11}\ll1$. Here, we calculate the  eigenvalues of $\hat{\mathcal{L}}$  perturbatively for $\epsilon\ll 1$. At $\epsilon=0$  the eigenvalues of $\hat{\mathcal{L}}$ coincides with the eigenvalues of $\mathcal{L}^{higher}$. 
At the  first order of the perturbative expansion, the eigenvalues and eigenvectors of $\hat{\mathcal{L}}$ are
\bea
&{\bm{v}} \approx \bm{v}_0+\epsilon \bm{v}',\\
&\hat{\lambda} \approx \lambda_0+\epsilon \lambda',
\eea
where, in order to preserve the unitary norm of $\bm{v}$ we need to impose, in the first order approximation, that $\bm{v}'$ is orthogonal to $\bm{v}_0$.
Substituting $\bm{v}$ and $\hat\lambda$ in  $\hat{\mathcal{L}}\bm{v}=\hat{\lambda}\bm{v}$ yields, in the first order approximation,
\begin{equation}\label{eig2}
    \mathcal{L}^{higher}\bm{v}'+ \mathcal{L}^{lower}\bm{v}_0=\lambda_0\bm{v}'+\lambda' \bm{v}_0.
\end{equation}
If $\lambda_0$ is non-degenerate eigenvalue (and hence $\lambda_0\neq 0$) the corresponding eigenvector of $\mathcal{L}^{higher}$ is  $\bm{v_0}=(\bm{u_0},\bm{u}_0)^{\top}$ where $\bm{u}_0$ is the eigenvector corresponding to the non degenerate eigenvalue $\lambda_0\neq 0$ of the multi-Laplacian matrix $L^{(1,1)}$.
By multiplying both sides of Eq. (\ref{eig2}) from the left by transpose of $\bm{v}_0$, we get
\begin{equation}
    \lambda'=\frac{\bm{v}_0^T\mathcal{L}^{lower}\bm{v}_0}{\bm{v}_0^T\bm{v}_0}=RQ\left(\frac{L^{[1]}+L^{[2]}}{2},\bm{u}_0\right).
\end{equation}
Hence, the eigenvalues $\lambda$ of Hyper-Laplacian that are perturbation of  non-zero and non-degenerate eigenvalues $\lambda_0$, are given for $\delta_{11}=1/\epsilon\gg1$ by 
\be\label{nonzerof divergent}
\lambda\approx\delta_{11}\hat{\lambda}= \delta_{11}\lambda_0+RQ\left(\frac{L^{[1]}+L^{[2]}}{2},\bm{u}_0\right).
\ee
Notice that since $\lambda_0>0$, all these eigenvalues scale like $O(\delta_{11})$ as $\delta_{11}\gg1$ and hence diverge with $\delta_{11}$. The asymptotic expansion of non-degenerate eigenvalues can be conducted following similar steps as in \ref{Ap:degenerate}.

If $\lambda_0=0$, let us consider $\bm{v}_0=\left( \bm{u}_0^{[1]},\bm{u}_0^{[2]}\right)^{\top}$ indicating the eigenvector of $\mathcal{L}^{higher}$ corresponding to eigenvalue $\lambda_0=0$ and a a vector $\bm{v}'=\left( \bm{u}'^{[1]},\bm{u}'^{[2]}\right)^{\top}$ perpendicular to the kernel fo $\mathcal{L}^{higher}$. In these hypothesis  Eq. (\ref{eig2}) admits the explicit expression
\bea
    L^{(1,1)}(\bm{u}'^{[1]}+\bm{u}'^{[2]})+L^{[1]} \bm{u}_0^{[1]}=\lambda' \bm{u}_0^{[1]},\label{firsteq}\\
    L^{(1,1)}(\bm{u}'^{[1]}+\bm{u}'^{[2]})+L^{[2]} \bm{u}_0^{[2]}=\lambda' \bm{u}_0^{[2]}.\label{secondeq}
\eea
By subtracting equations (\ref{firsteq}) and (\ref{secondeq}) we get
\begin{equation}\label{diffeq}
    L^{[1]}\bm{u}_0^{[1]}-L^{[2]}\bm{u}_0^{[2]}=\lambda'(\bm{u}_0^{[1]}-\bm{u}_0^{[2]}).
\end{equation}
In the following, we consider a particular case where multi-Laplacian $L^{(1,1)}$ has exactly one eigenvalue equal to zero; in other words, the case where network of overlap of two layers is connected. In this scenario, the eigenvector $\bm{v}_0$ corresponding to eigenvalue $\lambda_0=0$ of $\mathcal{L}^{higher}$ can be represented as
\bea
    \bm{v}_0=(\bar{\bm{u}},-\bar{\bm{u}})^{\top}+c(\bm{1},\bm{1})^{\top}
    \label{Gv01}
\eea
where $\bm{1}$ is all one $N$-dimensional vector whose elements are all equal to one, and $\bar{\bm{u}}$ is an  arbitrary $N$ dimensional vector. If we substitute  Eq.(\ref{Gv01}) into equation (\ref{diffeq}), we have:
\begin{equation}
    \left(\frac{L^{[1]}+L^{[2]}}{2}\right)\bar{\bm{u}}=\lambda'\bar{\bm{u}}.
\end{equation}
Hence, $\lambda'=\mu$ where $\mu$ is the eigenvalue of  matrix $(L^{[1]}+L^{[2]})/2$. It follows that the associated eigenvalues of Hyper-Laplacian $\lambda$ are
given, in the first order approximation, by  $\lambda=\delta_{11}(\lambda_0+\epsilon\mu)$, and since $\lambda_0=0, \delta_{11}\epsilon=1$, we obtain
\begin{equation}
    \lambda \approx \mu.
\end{equation}
Therefore, when multi-Laplacian $L^{(1,1)}$ has exactly one eigenvalue equal to zero, the smallest non-zero eigenvalue of Hyper-Laplacian $\mathcal{L}$ approaches its upper bound for $\delta_{11}\rightarrow{\infty}$.
It follows that in the limit $\delta_{11}\gg1$ as long as the network formed by multilinks of type $(1,1)$ is connected, the un-normalised Hyper-Laplacian  displays  $2N-2$ non-zero eigenvalues, of which $N-1$ eigenvalues  scale like $O(\delta_{11})$ for $\delta_{11}\gg1$  and hence are diverging with $\delta_{11}$ while $N-1$ eigenvalues remain $O(1)$ and  have a finite limit for $\delta_{11}\to \infty$.
Similarly on can show that under the same hypothesis the normalised Hyper-Laplacian $\mathcal{L}^{norm}$ displays exactly $2N-2$ non-zero eigenvalues of which $N-1$ eigenvalues are $O(1)$ and converge to a constant as $\delta_{11}\to \infty$ while the remaining non-zero $N-1$ eigenvalues scale as  $O(1/\delta_{11})$ for $\delta_{11}\gg 1$ (see \ref{Ap:dgg1}).

\begin{figure}[!htbp]
\begin{center}
$\hspace*{-15mm}\includegraphics[width=1.20\textwidth]{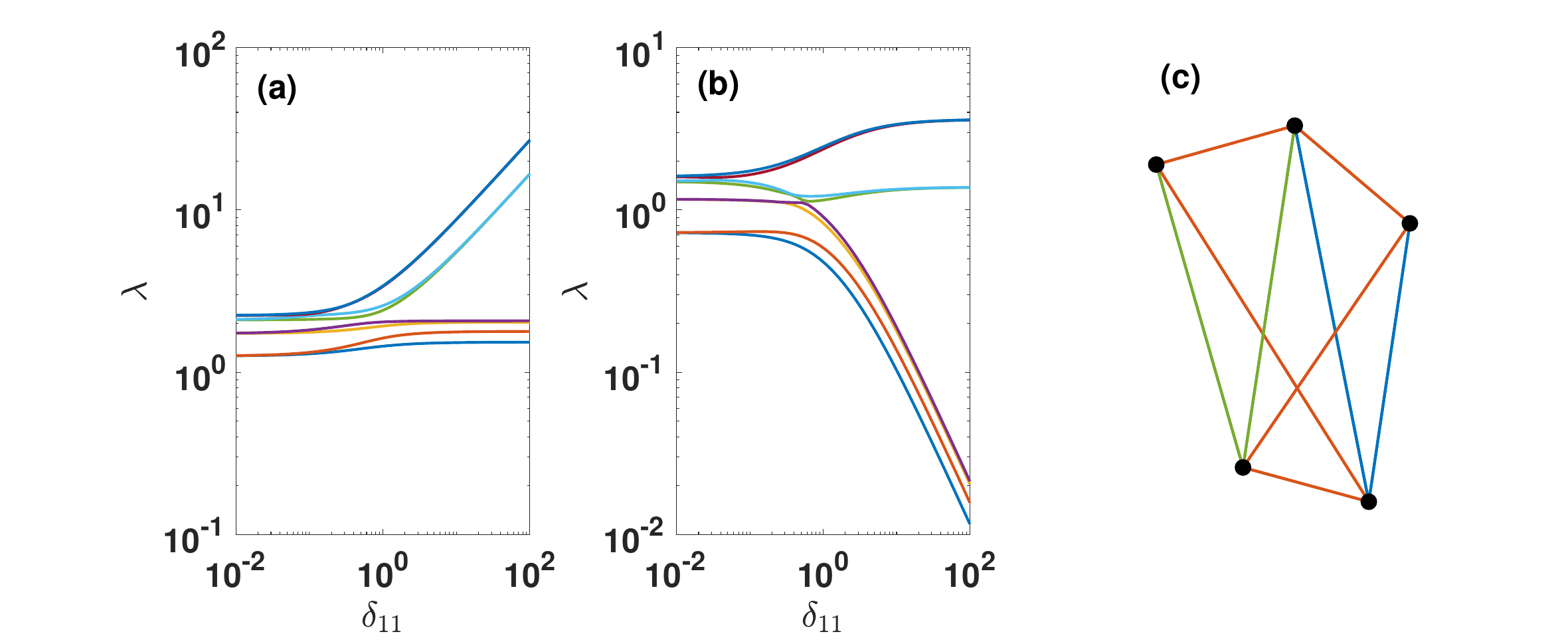}
$
\caption{{\bf The non-zero spectrum of the Hyper-Laplacian on a small  {duplex} network.}
The eight non-zero eigenvalues of a duplex network with $N=5$ nodes are plotted as a function of $\delta_{11}$ for the Hyper-Laplacian $\mathcal{L}$ (panel (a)) and the normalised Hyper-Laplacian $\mathcal{L}^{norm}$. The duplex network is shown in panel (c) with the multilinks of type $(1,0)$ in blue color, the multilinks $(0,1)$ in green, 
and the multilinks of type $(1,1)$ in red.}
\label{fig:5}
\end{center}
\end{figure}
\subsection{Example of the spectrum of a small network}

In order to provide a concrete example where the results of the above asymptotic expansion can be tested, we consider a small network of $N=5$ and we study the spectrum of the un-nonrmalised Hyper-Laplacian and the spectrum of the normalised Hyper-Laplacian as a function of $\delta_{11}$ (see Figure \ref{fig:5}).
The considered duplex network is formed by two connected layers, and the network formed by multilinks of type $(1,1)$ is connected, therefore as long as $\delta_{11}>0$  {the zero eigenvalue} has multiplicity $2$ and the number of non-zero eigenvalues is $8$.
For the un-normalised Hyper-Laplacian we observe that exactly $4$ eigenvalues diverge as $\delta_{11}\to \infty$ while the other eigenvalues have a finite limit. For the normalised Hyper-Laplacian we observe instead that $4$ eigenvalues  have a finite limit as $\delta_{11}\to \infty$ while four eigenvalues scale as $O(1/\delta_{11})$ as $\delta_{11}\gg1$ (see Figure $\ref{fig:5}$). Interestingly,  in the considered example we observe an eigenvalue crossing as a function of $\delta_{11}$ for the normalised Hyper-Laplacian.

\section{Conclusions}
 {In this work,  we have introduced multiplex hyper-diffusion,  
 {a novel type 
of diffusion on  multiplex networks that also takes into 
account higher-order interactions}, 
and we have investigated its dynamical properties by means of spectral graph theory.  {The main idea behind multiplex  hyperdiffusion is that it accounts for two processes at the same time:} diffusion occurring on each single layer of the multiplex network due to the specific  intra-layer pairwise interactions, and higher-order diffusion among different layers occurring when links of different layers overlap. 
Such higher-order interactions due to link overlap are here treated similarly to the oriented hyperedges studied in Refs.\cite{jost2019hypergraph,mulas2020coupled}.\\
In this work,  we have focused in particular on hyper-diffusion on duplex networks  {,i.e. multiplex networks with two layers}. In a duplex network hyper-diffusion couples the dynamics of the two layers of a multiplex network through multilinks of type $(1,1)$, which encode 4-body interactions.  The relevance  of  higher-order diffusion with respect to diffusion occurring in each  {of the two}
single layers  is controlled by the parameter $\delta_{11}$.}
We have investigated the properties of hyper-diffusion  {as a function of the value of the tuning parameter 
$\delta_{11}\ge 0$} by characterising the  spectral properties of the Hyper-Laplacians that drive this dynamical processes. Specifically, we have provided  upper and lower-bounds for the Fiedler eigenvalue and the largest eigenvalue of the Hyper-Laplacians, and we have compared  these theoretical bounds with simulation results.
In addition to this, we have also investigated the dynamical properties of hyper-diffusion and its relaxation to the steady state.
We found that hyper-diffusion does not lead to a ``transfer of mass" between the two layers, as the average state (or a weighted average state) of the replica nodes is a conserved quantity of the dynamics. However, the presence of higher-order interactions couples the dynamics in the two layers and allows the two layers to reach the steady state at the same time, as long as the eigenvector associated to the Fielder eigenvalue of the Hyper-Laplacian is not localised on a single layer. This implies that when, in absence of hyper-diffusion  {($\delta_{11}=0$)} 
the two layers would display different characteristic relaxation times, as long as the hyper-diffusion on multilinks of type $(1,1)$ is turned on 
 {($\delta_{11} > 0$)}, and the Fiedler eigenvector of the Hyper-Laplacian is delocalized in both layers, the two layers display the same relaxation time.
Finally ,{} we have investigated the asymptotic behaviour of hyper-diffusion in the limiting cases when $\delta_{11}\ll1$ and $\delta_{11}\gg 1$, highlighting the differences among hyper-diffusion driven by the un-normalised and by the asymmetric normalised Hyper-Laplacian.

In conclusion this is a pioneering work that treats diffusion on multiplex networks with  higher-order interactions, and shows that the  higher-order coupling between the layers changes the diffusion properties of the entire multiplex network.  Finally, our work establishes a bridge between the study of multiplex networks and the study of higher-order networks and oriented hypergraphs, and we hope that this can lead to further research at the interface of these two hot topics.
 Possible  directions in which this work can be expanded include the  investigation of hyper-diffusion on multiplex networks with more than two layers and the application of this framework to non-linear models, such as synchronisation of identical oscillators or generalised Kuramoto models.

\section*{Data availability.}
No real data have been used in this work.

\section*{Code availability.}
The code used to generate multiplex networks with given multidegree distribution is available at \cite{Git_repository}.

\section*{References}
\bibliographystyle{unsrt}
\bibliography{higherorderdiffusion}
\appendix

\section{Dimension of the kernel of the Hyper-Laplacians}
\label{Ap:kernel}
We consider a duplex network formed by two layers, each having $p^{[\alpha]}$ connected components with support on the sets of replica nodes $S_1^{[\alpha]},S_2^{[\alpha]}\ldots, S_{p^{[\alpha]}}^{[\alpha]}$.
It is well known that the dimension of the Laplacian $L^{[\alpha]}$ of layer $\alpha$ is $p^{[\alpha]}$ and it is spanned by the orthogonal $N$-dimensional vectors $\bm{x}^{n,[\alpha]}$ having value one on $S_n^{[\alpha]}$ and zero everywhere else.

From this well known result and from the fact that  $\mathcal{L}^{lower}$ is  the direct sum of $L^{[1]}$ and $L^{[2]}$, it follows that the dimension of the kernel of  $\mathcal{L}^{lower}$ is  $p=p^{[1]}+p^{[2]}$ and that this kernel  is spanned by the orthogonal $2N$-dimensional vectors $\bm{X}^{n,[1]}=(\bm{x}^{n,[1]},{\bf 0})^{\top},\bm{X}^{n,[2]}=({\bf 0},\bm{x}^{n,[2]})^{\top}$.

Here {, we }want to show that the dimension of the kernel  of the Hyper-Laplacian $\mathcal{L}=\mathcal{L}^{lower}+\delta_{11}\mathcal{L}^{higher}$  is also  $p$, and that the kernel of $\mathcal{L}$ coincides with the kernel of $\mathcal{L}^{lower}$, i.e.
\bea
\mbox{ker}(\mathcal{L})=\mbox{ker}\mathcal{L}^{lower}.
\eea

In order to prove this result,  first we observe that since $\mathcal{L}=\mathcal{L}^{lower}+\delta_{11}\mathcal{L}^{higher}$ and both Laplacians $\mathcal{L}^{lower}$ and $\mathcal{L}^{higher}$ are semipositive definite, 
\bea
\mbox{ker}(\mathcal{L})\subseteq \mbox{ker}(\mathcal{L}^{lower}).
\eea

Secondly we show that 
\bea
\mbox{ker}(\mathcal{L})\supseteq \mbox{ker}(\mathcal{L}^{lower}), 
\label{due}
\eea
by demonstrating that the vectors $\bm{X}^{n,[\alpha]}$ that  {span} the kernel of $\mathcal{L}^{lower}$ are also in the kernel of  $\mathcal{L}^{higher}$ and hence in the kernel of $\mathcal{L}$.
We have already mentioned that the vectors $\bm{x}^{n,[\alpha]}$ are in the kernel of $L^{[\alpha]}$, which implies
\bea
\langle \bm{x}^{n,[\alpha]},L^{[\alpha]}\bm{x}^{n,[\alpha]}\rangle=\frac{1}{2}\sum_{i,j}a_{ij}^{[\alpha]}\left(x^{n,[\alpha]}_i-x^{n,[\alpha]}_j\right)^2=0,
\label{A4}
\eea
where here and in the following we indicate the scalar product between the generic column vectors $\bm{u}$ and $\bm{w}$ as $\langle \bm{u},\bm{w} \rangle=\bm{u}^{\top}\bm{w}$.
Eq.(\ref{A4}) implies that $x^{n,[\alpha]}_i=x^{n,[\alpha]}_j$ whereas $a_{ij}^{[\alpha]}=1$.
We now want to show that $\bm{X}^{n,[\alpha]}$ are in the kernel of $\mathcal{L}^{higher}$ which implies
\bea
\langle \bm{X}^{n,[\alpha]},\mathcal{L}^{higher}\bm{X}^{n,[\alpha]}\rangle=\frac{1}{2}\sum_{i,j}A_{ij}^{(1,1)}\left(x^{n,[\alpha]}_i-x^{n,[\alpha]}_j\right)^2=0.
\eea
This relation is identically satisfied as whereas $A_{ij}^{(1,1)}=1$ then $a_{ij}^{[\alpha]}=A_{ij}^{(1,0)}+A_{ij}^{(1,1)}=1$, and therefore $x^{n,[\alpha]}_i-x^{n,[\alpha]}_j=0$. Hence,  we have demonstrated the relation $(\ref{due})$ completing the proof that 
\bea
\mbox{dim}\ \mbox{ker}(\mathcal{L})=p.
\eea

We conclude by noticing that also the dimension of the normalised Hyper-Laplacian $\mathcal{L}^{norm}$ is $p$.
Indeed the dimension of the kernel of the normalised Hyper-Laplacian is the dimension of the space of all vectors $\bm{X}$ for which 
 the Rayleigh quotient of the normalised Hyper-Laplacian vanishes, i.e.
 \bea\label{raylnorm}
    RQ(\mathcal{L}^{norm})=\frac{\bm{X}^T\mathcal{L}^{norm}\bm{X}}{\bm{X}^T\bm{X}}=0.\eea
     Since we have 
     \bea     
     RQ(\mathcal{L}^{norm})=\frac{\bm{X}^TD^{-\frac{1}{2}} \mathcal{L} D^{-\frac{1}{2}} \bm{X}}{\bm{X}^T \bm{X}}=\frac{\bm{Y}^T \mathcal{L} \bm{Y}}{\bm{Y}^T D \bm{Y}}
      \eea
 where $\bm{Y}=D^{-\frac{1}{2}}\bm{X}$, it follows that the dimension of the kernel of $\mathcal{L}^{norm}$ is equal to the dimension of the space formed by the vectors $\bm{Y}$ for which $\bm{Y}^T \mathcal{L} \bm{Y}=0$, i.e. the dimension of the kernel of $\mathcal{L}$,
or equivalently 
\bea
\mbox{dim}\ \mbox{ker}(\mathcal{L}^{norm})=p.
\eea

\section{Proof of Eq. (\ref{gen1}) and Eq. (\ref{gen2})}
\label{Ap:gen}
In this Appendix we provide the proof of Eq.(\ref{gen1}) and Eq. (\ref{gen2}).
Let us first prove Eq. (\ref{gen1}) for the eigenvalues of $\mathcal{L}$ and then generalise the results for the eigenvalues of $\mathcal{L}^{norm}$ (Eq.(\ref{gen2}).
Assume that the zero eigenvalue of $\mathcal{L}$ has degeneracy $s-1$. 
We indicate with  $\lambda_s(\mathcal{L})$ be the smallest non-zero eigenvalue  and with $\lambda_L(\mathcal{L})$ be the largest eigenvalue of $\mathcal{L}$. 
Then it follows that 
\bea
\hspace*{-15mm}(2N-(s-1))\lambda_s(\mathcal{L}) \leq \sum_{i=s}^{2N}\lambda_i = \sum_{i=1}^{2N}\lambda_i =\mbox{Tr}\mathcal{L} \leq (2N-(s-1)) \lambda_L(\mathcal{L}),
  \eea
This relation yields
\bea
  \lambda_s\left(\mathcal{L}\right) \leq \frac{\mbox{Tr}\mathcal{L}}{2N-(s-1)} \leq \lambda_L\left(\mathcal{L}\right).
\eea
Since $\mbox{Tr}\mathcal{L}=\sum_{i=1}^N(d_i^{[1]}+d_i^{[2]})$,  Eq. (\ref{gen1}) follows immediately, i.e.
\bea
  \lambda_s\left(\mathcal{L}\right) \leq \frac{\sum_{i=1}^N(d_i^{[1]}+d_i^{[2]})}{2N-(s-1)} \leq \lambda_L\left(\mathcal{L}\right).
\eea
Similarly, assuming that the zero eigenvalue of $\mathcal{L}^{norm}$ has degeneracy $s-1$, the following bounds hold
\bea
\lambda_s\left(\mathcal{L}^{norm}\right) \leq \frac{\mbox{Tr}(\mathcal{L}^{norm})}{2N-(s-1)} \leq \lambda_L\left(\mathcal{L}^{norm}\right).
\eea
  Since in a duplex network  $\mbox{Tr}(\mathcal{L}^{norm})\leq  2N$ with equality  if and only if there are no isolated nodes,  Eq.(\ref{gen2}) holds, i.e.
  \bea
  \lambda_s\left(\mathcal{L}^{norm}\right) \leq \frac{N}{N-(s-1)/2} \leq \lambda_L\left(\mathcal{L}^{norm}\right).
  \eea

\section{Proof of Eq.(\ref{ineq:lambda3})}
\label{Ap:l3}
Let us consider a duplex network formed by two connected layers. In this section we prove the inequalities of Eq. (\ref{ineq:lambda3}) which constitute upper and lower bounds for the Fiedler eigenvalue of $\mathcal{L}$.
 According to the Courant–Fischer theorem we know that the Fiedler eigenvalue of the considered duplex network can be obtained from the following minimization problem
\bea
\lambda_3(\mathcal{L})=\min_{\bm{X}\in Q}\bm{X}^{\top}\mathcal{L}\bm{X},
\label{c1}
\eea
where every vector $\bm{X}= \left(
\bm{x}^{[1]}, \bm{x}^{[2]}
\right)^{\top} \in Q$ should have the following properties:
\begin{itemize}
    \item $\bm{X}$ has to be normalised, i.e. $\bm{X}^{\top}\bm{X}=1$.
    \item $\bm{X}$ has to be perpendicular to the eigenspace of $\lambda =0 $ which implies: 
\bea    
\langle{\bm{X}},(\bm{1},\bm{0})^{\top}\rangle=0 \quad \langle\bm{X},(\bm{0},\bm{1})^{\top}\rangle=0,
\label{conditions}
\eea
\end{itemize}
where $\bm{0}$ and $\bm{1}$ are all zero and all one $N$-dimensional  vectors. 
To find a lower bound for the Fielder eigenvalue of $\mathcal{L}$ we observe that we can express  $\bm{X}^{\top}\mathcal{L}\bm{X}$ as \bea
\bm{X}^{\top}\mathcal{L}\bm{X}=\bm{X}^{\top}(\mathcal{L}^{lower}+\delta_{11}\mathcal{L}^{higher})\bm{X}.
\eea
Since $\mathcal{L}^{higher}$ is a semi positive definite operator, we have that 
\bea
\bm{X}^{\top}\mathcal{L}\bm{X}\geq \bm{X}^{\top}\mathcal{L}^{lower}\bm{X},
\eea
from which  it follows that 
\bea
\lambda_3(\mathcal{L})\geq \lambda_3(\mathcal{L}^{lower})=\min\left(\lambda_2(L^{[1]}),\lambda_2(L^{[2]})\right).
\eea
Therefore the Fiedler eigenvalue of $\mathcal{L}^{lower}$ is an lower-bound to the Fiedler eigenvalue of the Hyper-Laplacian $\mathcal{L}$.
Let us now find an upper bound for $\lambda_3(\mathcal{L})$. By substituting $\bm{X} =\left(\bm{x}^{[1]}, \bm{x}^{[2]}
\right)^{\top}$, the conditions in Eq.(\ref{conditions}) will be translated to:
$$\langle\bm{x}^{[1]},{\bm{1}}\rangle=0 \mbox{ \quad , } \langle\bm{x}^{[2]},\bm{1}\rangle=0 \mbox{ \quad , } \langle\bm{x}^{[1]},\bm{x}^{[1]}\rangle+\langle\bm{x}^{[2]},\bm{x}^{[2]}\rangle=1. $$ Moreover we can express Eq. (\ref{c1}) in terms of the vectors $\bm{x}^{[1]}$ and $\bm{x}^{[2]}$ as
\begin{equation}
\lambda_s(\mathcal{L})=\min_{\bm{X}\in Q}{F}({\bm{x}}^{[1]},{\bm{x}}^{[2]}) 
\end{equation}
where ${F}({\bm{x}}^{[1]},{\bm{x}}^{[2]})$ is given by 
\bea
{F}({\bm{x}}^{[1]},{\bm{x}}^{[2]})&=&(\bm{x}^{[1]})^TL^{[1]} \bm{x}^{[1]}+(\bm{x}^{[2]})^TL^{[2]}\bm{x}^{[2]}+2 (\bm{x}^{[1]})^TL^{(1,1)}\bm{x}^{[2]}\nonumber \\
&&+(\bm{x}^{[1]})^TL^{(1,1)}\bm{x}^{[1]}+ (\bm{x}^{[2]})^TL^{(1,1)}\bm{x}^{[2]} .
\eea
If we  restrict the domain in which we are taking the minimum to vectors  $\bm{X} =\left(\bm{x}^{[1]}, \bm{x}^{[2]}\right)$ with $\langle\bm{X},\bm{X}\rangle=1$, $\bm{x}^{[1]}=-\bm{x}^{[2]}=\bm{u}$ and $\langle\bm{u},\bm{1}\rangle=0$ , we obtain an upper bound for the Fiedler eigenvalue given by 
\begin{equation}
\lambda_s(\mathcal{L})\leq \min_{\bm{u}}\bm{u}^T(L^{[1]}+L^{[2]})\bm{u}.
\end{equation}
where
$\langle\bm{u},\bm{\bm{1}}\rangle=0$ , $\langle\bm{u},\bm{u}\rangle=\frac{1}{2}$. 
Considering the normalised $N$-dimensional vector $\bm{u}'=\sqrt{2}\bm{u}$ we obtain
\begin{equation*}
\lambda_s(\mathcal{L})\leq \min_{\bm{u}'}\Big\{(\bm{u}')^T\frac{(L^{[1]}+L^{[2]})}{2}\bm{u}'\Big\}=\lambda_s\left(\frac{L^{[1]}+L^{[2]}}{2}\right)
\end{equation*}
where $\langle\bm{u}',1\rangle=0$ , $\langle\bm{u}',\bm{u}'\rangle=1$.\\
Hence, the upper bound for the Fiedler eigenvalue of Hyper-Laplacian $\mathcal{L}$ is the Fiedler eigenvalue of the weighted aggregated network of the duplex network.

\section{Proof of Eq. (\ref{lambdaL}) and Eq. (\ref{lambdaL2})}
\label{Ap:lambdaL}
In this Appendix we proof Eq.(\ref{lambdaL}) and Eq.(\ref{lambdaL2}) providing an upper bound for the largest eigenvalue $\lambda_L(\mathcal{L})$ and  $\lambda_L(\mathcal{L}^{norm})$ respectively.
Let us start by proving Eq. (\ref{lambdaL}) providing the upper bound for the largest eigenvalue  $\lambda_L(\mathcal{L})$ of the non-normalised Hyper-Laplacian $\mathcal{L}$.

It is a well known graph theory result that the largest eigenvalue of a Laplacian matrix, can be found by maximising the Rayleigh quotient. In our case we have therefore 
\bea
\lambda_L\left(\mathcal{L}\right)= \max_{\bm{X}\in \mathbb{R}^{2N}}\frac{\bm{X}^{\top}\mathcal{L}\bm{X}}{\bm{X}^{\top}\bm{X}},
\eea

where, setting $\bm{X}=(\bm{x}^{[1]},\bm{x}^{[2]})^{\top}$, the explicit expression of $\bm{X}^{\top}\mathcal{L}\bm{X}$ is given by 
\bea
\bm{X}^{\top}\mathcal{L}\bm{X}&=&\frac{1}{2}\sum_{i,j}a^{[1]}_{ij}(x^{[1]}_i-x^{[1]}_j)^2
     +\frac{1}{2}\sum_{i,j}a^{[2]}_{ij}(x^{[2]}_i-x^{[2]}_j)^2\nonumber \\
     && + \frac{1}{2}\delta_{11}\sum_{i,j}A^{(1,1)}_{ij}(x^{[1]}_i-x^{[1]}_j+x^{[2]}_i-x^{[2]}_j)^2.
\eea
Using the inequality $(a\pm b)^2\leq 2(a^2+b^2)$ twice we obtain
\bea
\bm{X}^{\top}\mathcal{L}\bm{X}&\leq& \sum_{i,j}a^{[1]}_{ij}\left[(x^{[1]}_i)^2+(x^{[1]}_j)^2\right]
     +\sum_{i,j}a^{[2]}_{ij}\left[(x^{[2]}_i)^2+(x^{[2]}_j)^2\right]\nonumber \\
     && + 2\delta_{11}\sum_{i,j}A^{(1,1)}_{ij}\left[(x^{[1]}_i)^2+(x^{[1]}_j)^2+(x^{[2]}_i)^2+(x^{[2]}_j)^2\right].
     \eea
     Note that here the equality holds if and only if $x_i^{[\alpha]}=-x_j^{[\alpha]}$ whereas $a_{ij}^{[\alpha]}=1$ and $x_i^{[1]}=x_i^{[2]}=-x_j^{[1]}=-x_j^{[2]}$ whereas $A_{ij}^{(1,1)}=1$ which is a configuration only possible when both layers are bipartite.
     Using the fact that the network are undirected, and therefore the adjacency matrices $a^{[1]}$ and $a^{[2]}$ are symmetric, and the multiadjacency matrix $A^{(1,1)}$ is symmetric as well we obtain
     \bea
 \bm{X}^{\top}\mathcal{L}\bm{X}&\leq & 2\sum_{i}d_i^{[1]}(x_i^{[1]})^2+2\sum_{i}d_i^{[2]}(x_i^{[2]})^2+2\delta_{11}\sum_{i}k_i^{(1,1)}(x_i^{[1]})^2\nonumber \\&&+2\delta_{11}\sum_{i}k_i^{(1,1)}(x_i^{[2]})^2.
 \label{d14}
\eea
Using the following relation between the hyper-degrees and the multidegrees, 
\bea
d_i^{[1]}=k^{(1,0)}_i+(1+\delta_{11})k^{(1,1)}_i,\quad
d_i^{[2]}=k^{(0,1)}_i+(1+\delta_{11})k^{(1,1)}_i,
\eea
it is immediate to show that 
\bea
k_i^{(1,1)}\leq d_i^{[1]}\frac{1}{1+\delta_{11}},\quad
k_i^{(1,1)}\leq d_i^{[2]}\frac{1}{1+\delta_{11}}.
\eea
Note that here the equality holds for every node $i$ only if the two layers are identical and the duplex network is only formed by multilinks of type $(1,1)$.
Inserting this inequality in Eq. (\ref{d14}) we obtain 
\bea
 \bm{X}^{\top}\mathcal{L}\bm{X}&\leq& \left(2+2\frac{\delta_{11}}{1+\delta_{11}}\right)\bm{X}^{\top}D\bm{X},
 \label{d15}
\eea
where
\bea
\bm{X}^{\top}D\bm{X}=\left[\sum_{i}d_i^{[1]}(x_i^{[1]})^2+\sum_{i}d_i^{[2]}(x_i^{[2]})^2\right]\leq d_{max}\bm{X}^{\top}\bm{X}.
\label{d16}
\eea
with
$d_{max}=\max\left(\max_{i=1,\ldots, N} d_i^{[1]},\max_{i=1,\ldots, N} d_i^{[2]}\right)$.
It follows that Eq. (\ref{lambdaL}) holds, i.e.
\bea
\lambda_L(\mathcal{L})=\max_{\bm{X}\in \mathbb{R}^{2N}} \frac{ \bm{X}^{\top}\mathcal{L}\bm{X}}{ \bm{X}^{\top}\bm{X}}\leq \left(2+2\frac{\delta_{11}}{1+\delta_{11}}\right) d_{max}.
\eea
The proof of Eq. (\ref{lambdaL2}) is a simple corollary of the previous result.
In fact the  largest eigenvalue $\lambda_L\left(\mathcal{L}^{norm}\right) $of the normalised Hyper-Laplacian is given by 
\bea
\lambda_L\left(\mathcal{L}^{norm}\right)= \max_{\bm{X}\in \mathbb{R}^{2N}}\frac{\bm{Y}^{\top}\mathcal{L}\bm{Y}}{\bm{Y}^{\top}D\bm{Y}}.
\eea
Using Eq.(\ref{d15}) we obtain that Eq.(\ref{lambdaL2}) holds, i.e.
\bea
\lambda_L(\mathcal{L}^{norm})=\max_{\bm{X}\in \mathbb{R}^{2N}} \frac{ \bm{Y}^{\top}\mathcal{L}\bm{Y}}{ \bm{Y}^{\top}D\bm{Y}}\leq \left(2+2\frac{\delta_{11}}{1+\delta_{11}}\right).
\eea

\section{Proof of Eq.(\ref{lambda_3_norm_upper})}
\label{Ap:upper2}
In this Appendix our goal is to prove Eq.(\ref{lambda_3_norm_upper}) providing an upper bound for the Fiedler eigenvalue of $\mathcal{L}^{norm}$ of a duplex networks formed by connected layers.
In particular we want to prove that is $\lambda_2(L^{[1]}_{norm})<\lambda_2(L^{[2]}_{norm})$ and if 
 \bea
  \frac{\bm{u}_1^{\top}L^{(1,1)}\bm{u}_1}{\bm{u}_1^{\top}D^{(1,1)}\bm{u}_1}\leq  \lambda_2(L^{[1]}_{norm}),
  \eea
  where $\bm{u}_1=[D^{[1]}]^{-\frac{1}{2}}\bm{y}_1$  with $\bm{y}_1$ indicating the eigenvector corresponding to the Fiedler eigenvalue of $L^{[1]}_{norm}$,
  then we haven 
\bea
\lambda_3(\mathcal{L}^{norm})\leq \lambda_2(L^{[1]}_{norm}).
\eea
 According to the Courant–Fischer theorem we know that the Fiedler eigenvalue $\lambda_3(\mathcal{L}^{norm})$ can be obtained from the following:
\bea
\lambda_3(\mathcal{L}^{norm})=\min_{\bm{X}\in Q}\frac{\bm{X}^{\top}\mathcal{L}\bm{X}}{\bm{X}D\bm{X}}.
\label{c12}
\eea
where $\bm{X}\in Q$ should have the following properties:
\begin{itemize}
    \item $\bm{X}$ has to be normalised, i.e. $\bm{X}^{\top}\bm{X}=1$.
    \item $\bm{X}$ has to be perpendicular to the kernel of $\mathcal{L}$ which means: 
\bea    
\langle{\bm{X}},(\bm{1},\bm{0})^{\top}\rangle=0 \quad \langle\bm{X},(\bm{0},\bm{1})^{\top}\rangle=0.
\label{conditions2}
\eea
\end{itemize}
Let us indicate with $\bm{u}_1=D_1^{\frac{1}{2}}{\bm{y}}_1$ where $\bm{y}_1$ is the Fiedler eigenvector of the normalised Laplacian $L^{[1]}_{norm}$, with $\langle \bm{u}_1,{\bf 1}\rangle=0$ and let us consider the vector $\bm{X}_1=(\bm{u}_1,{\bf 0})^{\top}$, then
\bea
\lambda_3(\mathcal{L}^{norm})\leq \frac{\bm{X}_1^{\top}\mathcal{L}\bm{X}_1}{\bm{X}_1D\bm{X}_1}.
\eea
We have that 
\bea
\bm{X}_1^{\top}\mathcal{L}\bm{X}_1=\bm{u}_1^{\top}[L^{(0,1)}+(1+\delta_{11})L^{(1,1)}]\bm{u}_1,\nonumber \\
\bm{X}_1D\bm{X}_1=\bm{u}_1^{\top}[D^{(1,0)}+(1+\delta_{11})D^{(1,1)}]\bm{u}_1.
\eea
Since $\bm{u}_1^{\top}L^{[1]}\bm{u}_1=\lambda_2(L^{[1]}_{norm})\bm{u}_1^{\top}D^{[1]}\bm{u}_1$ where $D^{[1]}=D^{(0,1)}+D^{(1,1)}$, we have 
\bea
\bm{X}_1^{\top}\mathcal{L}\bm{X}_1=\lambda_2(L^{[1]}_{norm})\bm{u}_1^{\top}[D^{(0,1)}+D^{(1,1)}]\bm{u}_1+\delta_{11}\bm{u}_1^{\top}L^{(1,1)}\bm{u}_1.
\label{u2}
\eea
Under our hypothesis we have
 \bea
\bm{u}_1^{\top}L^{(1,1)}\bm{u}_1\leq \lambda_2(L^{[1]}_{norm})[\bm{u}_1^{\top}D^{(1,1)}\bm{u}_1].\eea
Inserting this inequality in Eq.(\ref{u2}), using $D=D^{(0,1)}+(1+\delta_{11})D^{(1,1)}$ we obtain
\bea
\bm{X}_1^{\top}\mathcal{L}\bm{X}_1\leq \lambda_2(L^{[1]}_{norm})[\bm{X}_1D\bm{X}_1],
\eea
which implies 
\bea
\lambda_3(\mathcal{L}^{norm})=\min_{\bm{X}\in Q}\frac{\bm{X}^{\top}\mathcal{L}\bm{X}}{\bm{X}D\bm{X}}\leq \lambda_2(L^{[1]}_{norm}).
\label{c123}
\eea

\section{Proof of Eq. (\ref{lambdaL_norm_upper})}
\label{Ap:lambdaL2}

In this Appendix we proof Eq.(\ref{lambdaL_norm_upper}) using an approach that builds on the bounds provided in Ref.\cite{li2014bounds} and Ref.\cite{grone1994laplacian} for the largest eigenvalue of a generic graph Laplacian of a single network.\\
 The largest eigenvalue $ \lambda_L(\mathcal{L}^{norm})$ of the normalised Hyper-Laplacian $\mathcal{L}^{norm}$ can be found by maximising the Rayleigh quotient in equation, i.e.
  \begin{equation}
     \lambda_L(\mathcal{L}^{norm})=\max_{\bm{Y}\in \mathbb{R}^{2N}} 
     RQ(\mathcal{L}^{norm})=\max_{\bm{Y}\in \mathbb{R}^{2N}} \frac{\bm{Y}^T \mathcal{L} \bm{Y}}{\bm{Y}^T D \bm{Y}},
     \label{raylb}
  \end{equation}
  We split the proof of Eq.(\ref{lambdaL_norm_upper}) in two parts.
 \begin{itemize}
  \item 
  First our goal is to prove that the largest eigenvalue of the normalised Hyper-Laplacian satisfies
  \bea
    \lambda_L(\mathcal{L}^{norm})\geq \max{ \Bigg\{\frac{\Avg{d^{[1]}}N}{\Avg{d^{[1]}}N-d^{[1]}_{max}},\frac{\Avg{d^{[2]}}N}{\Avg{d^{[2]}}N-d^{[2]}_{max}} \Bigg\}}.
    \label{duno}
  \eea 
  where  $\Avg{d^{[\alpha]}}N=\sum_jd_j^{[\alpha]}$ and $d_{max}$ indicates the largest hyperdegree of all the replica nodes of the duplex network, i.e.
  $d_{max}=\max\left(\max_{i=1\ldots, N}d_i^{[1]},\max_{i=1\ldots, N}d_i^{[2]}\right)$.
 Consider the vector $\bm{Y}=\left(\bm{y}^{[1]},\bm{y}^{[2]}\right)^{\top}$  with $\bm{y}^{[\alpha]}$ given by 
   \bea
 y^{[1]}_j=
    \left\{ \begin{array}{ll}
        \Avg{d^{[1]}}N-d^{[1]}_i & \mbox{if}\  j=i,\\
        -d^{[1]}_i & \mbox{if}\  j\neq i,\\
      \end{array}\right.\quad  \quad y^{[2]}_j=0, \hspace{.5cm} \forall j,
 \eea
where $d^{[1]}_i=k^{[1]}_i+k^{(1,1)}_i$ is the  hyper-degree of an arbitrary node $i$ in the first layer. By substituting $\bm{Y}$ into the Eq. (\ref{raylb}), we have:
 \begin{equation}
 \hspace{-20mm}\lambda_L(\mathcal{L}^{norm})  \geq \frac{( \Avg{d^{[1]}}N)^2d^{[1]}_i}{( \Avg{d^{[1]}}N-d^{[1]}_i)^2 d^{[1]}_i+(d^{[1]}_i)^2(\Avg{d^{[1]}}N-d^{[1]}_i)}=
 \frac{\Avg{d^{[1]}}N}{\Avg{d^{[1]}}N-d^{[1]}_i}.
\end{equation}
Let $d^{[1]}_{max}$  be the largest duplex hyper-degree $d^{[1]}_i$ in layer $1$, i.e. $d^{[1]}_{max}=\max_{i=1\ldots, N}d_i^{[1]}$, then
\begin{equation}
\label{d1}
    \lambda_L(\mathcal{L}^{norm})  \geq \frac{\Avg{d^{[1]}}N}{\Avg{d^{[1]}}N-d^{[1]}_{max}}.
\end{equation}
In a similar fashion, we can obtain
\begin{equation}
    \lambda_L(\mathcal{L}^{norm})  \geq \frac{\Avg{d^{[2]}}N}{\Avg{d^{[2]}}N-d^{[2]}_{max}},
    \label{d2}
\end{equation}
where $d^{[2]}_{max}$ is the largest duplex hyper-degree in layer $2$, i.e. 
$d^{[2]}_{max}=\max_{i=1\ldots, N}d_i^{[2]}$.
Hence combining Eq. (\ref{d1}) and Eq.(\ref{d2}) it is straightforward to obtain Eq.(\ref{duno}).

\item 
Secondly our goal is to prove the following lower bound for the largest eigenvalue $\lambda_L(\mathcal{L}^{norm})$ of the normalised Hyper-Laplacian $\mathcal{L}^{norm}$:
\bea
\lambda_L(\mathcal{L}^{norm})\geq 1+\frac{1}{d_{max}}.
\label{m}
\eea
Let us indicate with $i$  the replica node in layer $1$ with the largest  hyper-degree  $d_{max}^{[1]}$.
We  consider the vector $\bm{Y}=\left(\bm{y}^{[1]},\bm{y}^{[2]}\right)^{\top}$ of elements
 \bea
 &y^{[1]}_j=\left\{
      \begin{array}{ll}
        d_{max}^{[1]} & \mbox{ if }\  j=i,\\
        -1 & \mbox{if}\  a_{ij}^{[\alpha]}=1, \\
        0 & \mbox{ otherwise},\\
      \end{array}\right.\quad \quad y^{[2]}_j=0 \  \forall j.
 \eea
The term $\bm{Y}^{\top}\mathcal{L}\bm{Y}$ can be bounded by considering only the contribution due to the interaction between node $i$ and its neighbours, 
\bea
\bm{Y}^{\top}\mathcal{L}\bm{Y}\geq \sum_j\left[a_{ij}^{[1]}+\delta_{11}A_{ij}^{(1,1)}\right](y_i^{[1]}-y_j^{[1]})^2=(d_{max}^{[1]}+1)^2 d_{max}^{[1]}.
\eea
Moreover{,} the term 
\bea
\bm{Y}^{\top}D\bm{Y}=d_{max}^{[1]}(d_{max}^{[1]})^2+\sum_{j}a_{ij}^{[1]}d_j^{[1]}
\eea
can be bounded by using $d_j^{[1]}\leq d_{max}^{[1]}$ with 
\bea
\bm{Y}^{\top}D\bm{Y}\leq  (d_{max}^{[1]})^2(d_{max}^{[1]}+1).
\eea
Therefore the maximum eigenvalue of the normalised Hyper-Laplacian satisfies 
 \bea
 \lambda_L(\mathcal{L}^{norm}) \geq \frac{\bm{Y}^{\top}\mathcal{L}\bm{Y}}{\bm{Y}^{\top}D\bm{Y}}
 &\geq 1+\frac{1}{d_{max}^{[1]}}.
 \label{m1}
 \eea
Similarly we can obtain 
\bea
\lambda_L(\mathcal{L}^{norm}) \geq 1+\frac{1}{d_{max}^{[2]}}.\label{m2}\eea
Therefore combining Eq. (\ref{m1}) and Eq.(\ref{m2}) we obtain Eq.(\ref{m})

\end{itemize}
\section{Asymptotic expansion of degenerate eigenvalues}
\label{Ap:degenerate}
Let us consider the asymptotic expansion of the eigenvalue of $\mathcal{L}$ for $\delta_{11}\ll1$ starting form a degenerate state.
We assume that at the first order of the asymptotic expansion the eigenvector $\bm{v}$ of $\mathcal{L} $ and its corresponding eigenvalue $\lambda$ are given by  Eq. (\ref{per_ll}) and satisfy Eq. (\ref{eig}) which we rewrite here for convenience, 
 \begin{equation}\label{eig_AP}
    \mathcal{L}^{lower}\bm{v}'+ \mathcal{L}^{higher}\bm{v}^{[\alpha]}=\lambda^{[\alpha]}\bm{v}'+\lambda'\bm{v}^{[\alpha]}.
\end{equation}
Here $\lambda^{[\alpha]}$ is a degenerate eigenvalue of $\mathcal{L}^{lower}$ and $\bm{v}^{[\alpha]}$ a generic eigenvector satisfying the characteristic eigenvalue Eq. (\ref{eig_alpha}).
Since the unperturbed eigenvalue $\lambda^{[\alpha]}$ is degenerate with degeneracy $g$, we consider a orthogonal basis of its eigenspace formed by eigenvalues $\bm{w}^{n}$ with $n=1,2,\ldots, g$  and we set 
\bea
\bm{v}^{[\alpha]}=\sum_{n'=1}^{g}c_{n'}\bm{w}^{n'}.
\eea
By inserting this expression in Eq.(\ref{eig_AP}) and multiplying both sides of Eq.(\ref{eig_AP}) from the left by $\bm{w}^{n}$ we obtain 
\bea
\sum_{n'=1}^{g}c_{n'}\langle\bm{w}^{n},\mathcal{L}^{higher}\bm{w}^{n'}\rangle=\lambda' c_{n}.
\eea
Finally by putting 
\bea
W_{n,n'}=\langle\bm{w}^{n},\mathcal{L}^{higher}\bm{w}^{n'}\rangle,
\eea
it is immediate to show that $\lambda'$ can be obtained by solving the  eigenvalue problem
\bea
\sum_{n'=1}^q W_{n,n'}c_{n'}=\lambda' c_n.
\eea
Therefore $\lambda'$ can be obtained by solving the characteristic equation 
\bea
\mbox{det}(W-\lambda' I)=0
\eea
where $I$ here indicates the $g\times g$ identity matrix.
\section{Asymptotic expansion of $\mathcal{L}^{norm}$ for $\delta_{11}\gg1$}
\label{Ap:dgg1}

The normalised Hyper-Laplacian $\mathcal{L}^{norm}$ is given by  \bea \mathcal{L}^{norm}=D^{-\frac{1}{2}}\mathcal{L}D^{-\frac{1}{2}}\eea where $\mathcal{L}=\mathcal{L}^{lower}+\delta_{11}\mathcal{L}^{higher}$ and 
\bea
D=\left(\begin{array}{cc}D^{[1]}+\delta_{11}D^{(1,1)}&0\\
0&D^{[2]}+\delta_{11}D^{(1,1)}
\end{array}\right).
\eea
For $\delta_{11}\gg 1$, the normalised Hyper-Laplacian
can be expanded as a function of $\epsilon=1/\delta_{11}\ll1 $ and in the first order approximation is given by 
\begin{equation}
    \mathcal{L}=\mathcal{L}^{higher}_{norm}+\epsilon \hat{\mathcal{L}}^{norm}
\end{equation}
where 
\bea
\hat{\mathcal{L}}^{norm}=\frac{1}{2}\{\mathcal{L}^{higher},\tilde{D}\}+\tilde{\mathcal{L}}
\eea
whit $\{,\}$ indicating the anti-commutator and  the matrices $\tilde{D}$ and $\tilde{L}$ indicating
\bea
\tilde{D}=\left(\begin{array}{cc}D^{[1]}[D^{(1,1)}]^{-1}&0\\
0&D^{[2]}[D^{(1,1)}]^{-1}\end{array}\right),\nonumber \\ \tilde{\mathcal{L}}=\left(\begin{array}{cc}[D^{(1,1)}]^{-\frac{1}{2}}L^{[1]}[D^{(1,1)}]^{-\frac{1}{2}}&0\\
0&[D^{(1,1)}]^{-\frac{1}{2}}L^{[2]}[D^{(1,1)}]^{-\frac{1}{2}}\end{array}\right).
\eea
Hence, the eigenvalues $\lambda$ of Hyper-Laplacian that are perturbation of  non-zero and non-degenerate eigenvalues $\lambda_0$ of $\mathcal{L}^{higher}_{norm}$, corresponding to the eigenvector $\bm{v}_0,$ are given for $\delta_{11}=1/\epsilon\gg1$ by 
\be\label{nonzerofdivergentAP}
\lambda\approx \lambda_0+\epsilon\frac{\bm{v}_0^T\hat{\mathcal{L}}^{norm}\bm{v}_0}{\bm{v}_0^T\bm{v}_0}.
\ee
Notice that since $\lambda_0>0$, all these eigenvalues have a finite limit as  $\delta_{11}\to \infty$.

If $\lambda_0=0$, let us consider $\bm{v}_0$ indicating the eigenvector of $\mathcal{L}^{higher}_{norm}$ corresponding to eigenvalue $\lambda_0=0$.
We consider the particular case in which the multi-Laplacian $L^{(1,1)}$ has exactly one zero eigenvalue or equivalently the case in which the network of multilinks of type $(1,1)$ is connected. In this scenario, the eigenvector $\bm{v}_0$ corresponding to eigenvalue $\lambda_0=0$ of $\mathcal{L}^{higher}$ can be represented as
\bea
    \bm{v}_0=(\bar{\bm{u}},-\bar{\bm{u}})^{\top}+c(\bm{1},\bm{1})^{\top}
    \label{Gv0}
\eea
where $\bm{1}$ is all one $N$-dimensional vector whose elements are all equal to one, and $\bar{\bm{u}}$ is an  arbitrary $N$ dimensional vector. By proceeding as in the case of asymptotic expansion of the un-normalised Hyper-Laplacian, we obtain  that to the first order approximation, the zero eigenvalue has degeneracy two and the perturbed  eigenvalues $\lambda$  of Hyper-Laplacian $\mathcal{L}^{norm}$ with lifted degeneracy are given by 
\begin{equation}
    \lambda \approx \frac{1}{\delta_{11}}\mu,
\end{equation}
where $\mu$ is the eigenvalue of  matrix $[D^{(1,1)}]^{-\frac{1}{2}}(L^{[1]}+L^{[2]})[D^{(1,1)}]^{-\frac{1}{2}}/{2}$ satisfying the eigenvalue problem
\begin{equation}
    [D^{(1,1)}]^{-\frac{1}{2}}\left(\frac{L^{[1]}+L^{[2]}}{2}\right)[D^{(1,1)}]^{-\frac{1}{2}}\bar{\bm{u}}=\lambda'\bar{\bm{u}}.
\end{equation}
Therefore these eigenvalue scale like $O(1/\delta_{11})$ for $\delta_{11}\gg1$.
\end{document}